\documentclass[12pt]{article}
\usepackage[top=0.75in, bottom=1in, left=1in, right=1in]{geometry}
\usepackage{graphicx}
\usepackage{placeins}
\usepackage{amsmath}
\usepackage{amssymb}
\usepackage{booktabs}
\usepackage[numbers,sort&compress]{natbib}
\usepackage{hyperref}
\usepackage{xcolor}
\usepackage{lineno}
\usepackage{setspace}
\usepackage{microtype}

\hypersetup{
  colorlinks=true,
  linkcolor=blue,
  citecolor=blue,
  urlcolor=blue
}

\title{\large\textbf{Going Beyond the d-band Center to Design Intermetallic
Catalysts for Nitrogen Reduction: A High-Throughput DFT and Machine Learning Study}}

\author{Parastoo Agharezaei$^{a}$,\ \ Kulbir Kaur Ghuman$^{a}$\thanks{Corresponding author: \texttt{Kulbir.Ghuman@inrs.ca}}}

\date{%
  \normalsize
  $^{a}$Institut National de la Recherche Scientifique, Centre \'{E}nergie
  Mat\'{e}riaux T\'{e}l\'{e}communications,\\
  1650 Boul.\ Lionel-Boulet, Varennes, Quebec, Canada J3X 1S2.
}

\begin{document}
\singlespacing
\maketitle
\vspace{-2em}
\begin{abstract}

This study combines high-throughput density functional theory (DFT) calculations
with machine learning (ML) techniques to uncover the key descriptors governing the
nitrogen reduction reaction (NRR) in intermetallic compounds (IMCs). A dataset of
47 bimetallic IMCs composed of Co, Ni, Al, Zn, V, Fe, Cu, and Pt was constructed,
and the adsorption energies of key intermediates (*N$_2$, *N$_2$H, and *NH$_3$)
were systematically evaluated across all accessible surface sites, yielding
approximately 1,200 data points. The analysis reveals that while *N$_2$ and
*N$_2$H adsorption energies remain strongly correlated, their scaling relationship
with *NH$_3$ is largely broken due to the presence of multimetallic adsorption
sites, enabling independent optimization of adsorption and desorption processes. Among the
IMCs studied, Fe$_9$Co$_7$ and Fe$_3$Co emerge as the most balanced catalysts,
exhibiting favorable adsorption energies across all three intermediates. By
incorporating intrinsic material properties along with local and global electronic
descriptors including s-, p- and d-band centers and fillings, as well as Bader
charges of atoms neighboring the adsorbate, predictive ML models were developed
with mean absolute errors (MAEs) of 0.27~eV for *N$_2$, 0.39~eV for *N$_2$H,
and 0.17~eV for *NH$_3$ adsorption. Importantly, accurate adsorption-energy
predictions were achieved using only 20 key features for *N$_2$H and *NH$_3$ and
38 features for *N$_2$, enabling the use of simple and computationally efficient
ML models. SHapley Additive exPlanations (SHAP) analysis indicates that p-band
and s-band characteristics play a more prominent role in determining adsorption
strength than the traditionally used d-band center, particularly for *N$_2$.
Beyond their established importance in systems containing p-block elements or
nearly filled d-band metals, s- and p-orbitals are also found to contribute
significantly to transition metal alloy activity such as Fe--Co, driven by
adsorption-induced sp--d hybridization. The influence of neighboring atoms is
shown to vary across intermediates, with *N$_2$ adsorption being particularly
sensitive to the local atomic environment. By challenging the d-band-centric
paradigm and identifying s- and p-band descriptors as critical yet overlooked
contributors, this work redefines the electronic descriptor space for
intermetallic NRR catalysts and lays the groundwork for DFT--ML-guided discovery
of non-noble, compositionally complex materials for sustainable ammonia
synthesis.

\medskip
\noindent\textbf{Keywords:} intermetallic compounds, machine learning, nitrogen
reduction reaction, electronic descriptors, catalyst design
\end{abstract}
\doublespacing
\newpage
\section{Introduction}

Ammonia is one of the central materials of modern industry and agriculture: it is an
essential chemical used for fertilizers that enable large-scale food production and is
also increasingly recognized as a clean, carbon-free energy carrier~\cite{Collado2024}.
Yet, producing it today still depends mainly on the Haber--Bosch process (HBP), a method
developed over a century ago that requires high temperature and pressure. This industrial
route consumes nearly 1--2\% of global energy and releases more than 450 million tons of
CO$_2$ every year~\cite{RafiqulBari2024,Zhang2024}. To address these concerns,
researchers have investigated several alternative routes for ammonia synthesis under mild
conditions, including photocatalytic, electrochemical, and biological nitrogen fixation,
or plasma-assisted catalytic
processes~\cite{Guan2024,Wang2018,Puertolas2022,Fu2021,Shao2023,Wen2021,Jin2020,Dixon2004,Li2023,Fritz2024,Iwamoto2017,Zhu2022}.
Despite their promise, these strategies are not yet practical replacements for HBP, as
their efficiency and scalability are below industrial standards. Consequently, designing
catalysts capable of synthesizing ammonia efficiently under milder conditions remains a
critical research priority. The main challenge in NRR is nitrogen activation. The
exceptionally strong N$\equiv$N bond makes its activation highly challenging, and the
competing hydrogen atoms poison the surface by occupying available adsorption sites on the
catalyst surface~\cite{Ren2022,Chang2024}. Even advanced catalysts, from noble metals such
as Ru, Pt and Au, to transition-metal compounds still struggle to achieve the necessary
balance between activity, selectivity, and long-term stability~\cite{Mehta2018}.

Catalysts based on transition metals (TMs) are attractive for NRR because their high
electron density and partially filled d-orbitals enable the well-known
``acceptance--donation'' mechanism, (accepting $\sigma$-electrons from N$_2$ and
back-donating into the $\pi^*$ orbitals of transition metals)~\cite{Chen2021}. Yet, proton
adsorption is usually more favorable than N$_2$ adsorption on TM sites, which suppresses
N$_2$ activation~\cite{Suryanto2019}. Moreover, linear scaling relations (LSRs) impose an
additional limitation, where catalysts that bind N$_2$ strongly often struggle with NH$_3$
desorption or suffer from hydrogen poisoning~\cite{Montoya2015}. To overcome these issues,
alloying has been proposed as an effective route to build multielement sites on the surface
of catalysts that combine the advantageous properties of different metals, enabling tuning
of the structural and electronic properties~\cite{Liu2023,Gong2018,Yin2021}. Indeed, alloy
catalysts such as solid solution alloys (e.g.\ high entropy alloys) or single-atom alloys
(SAAs) can partly bypass the LSRs by controlling the number of distinct neighboring atoms
of adsorption sites to achieve improved
performance~\cite{Agharezaei2025,Hutu2024,Sun2018}. However, the coexistence of
multiple types of adsorption sites in alloys that have intrinsic disorder often gives rise
to local scaling relations, which limit their effectiveness~\cite{Saidi2022}. Intermetallic
compounds (IMCs) have emerged as a group of materials for catalytic NRR because their
ordered crystal structure and uniform adsorption sites enable durable performance and
clearer structure--activity understanding compared to disordered
alloys~\cite{Zhou2023Ni3Mo,Wang2021}. In IMCs, the adsorption of *N$_2$ and
*N$_x$H$_x$ intermediates can be tuned to improve overall selectivity. The multielement
adsorption sites in IMCs enable the breaking of the LSRs by separating adsorption sites for N$_2$
activation versus *N$_x$H$_x$ hydrogenation~\cite{Liu2023,Gong2018,Yin2021,Wang2021,Wei2025,Fan2021}.
Furthermore, IMCs can be synthesized easily using phase diagrams~\cite{Zhou2023Ni3Mo}.

There are only a limited number of studies that have systematically screened the NRR
activity and selectivity of IMCs. Zhou et al.~\cite{Zhou2023ACS} investigated 29 L1$_2$
(A$_3$B) IMCs using high-throughput DFT calculations comprised of elements such as Ag, Cu,
Ta, Pd, Ru, and Mo. They analyzed how variations in composition of adsorption site and
electronic structure affect adsorption configurations and binding energies, and identified
Pd$_3$Mo as the most promising candidate, offering improved selectivity with a limiting
potential of $\approx -0.31$~V. Shamekhi et al.~\cite{Shamekhi2025} reported a combined
high-throughput DFT and machine learning (ML) study to identify efficient ordered bimetallic
alloy catalysts for the NRR. They first generated a DFT dataset of adsorption energies for
NRR key intermediates (*N$_2$, *N$_2$H, *NH, *NH$_2$) across 220 ordered structures and by
using electronic d-band features and intrinsic atomic properties as descriptors, they trained
an artificial neural network (ANN) capable of predicting the limiting potential ($U_L$) of
alloys with an accuracy of 0.23~eV, comparable to DFT. The model was then used to screen
other alloy surfaces, with Au@Au$_3$Re and Au@Au$_3$Mo identified as promising candidates.
Their study confirmed that alloying induces charge transfer from Re/Mo to Au, shifting
d-band centers and enhancing N$_2$ activation, while also improving selectivity by favoring
N$_2$ adsorption over H adsorption. In another study, Zhou et al.~\cite{Zhou2023Ni3Mo}
proposed Ni$_3$Mo as an NRR catalyst, combining the strong N$_2$ affinity of Mo with the
higher H affinity of Ni to separate adsorption sites for N$_2$ activation from *NH$_x$
hydrogenation, thereby bypassing LSRs and suppressing HER. This site separation is evidenced
by hydrogen migrating away from Mo top sites and by more favorable *NH/*NH$_2$ formation on
Ni$_3$Mo compared to pure Mo.

Despite extensive research on NRR catalysts, most studies have focused on single metals or
disordered alloys, leaving the structure-property relationships of ordered intermetallic
compounds (IMCs) largely unexplored. Existing DFT investigations typically consider only a
few representative surface sites, often one to three per material, and therefore overlook
how variations in the local electronic environment influence adsorption energies and NRR
activity. In this article, we first describe the construction of adsorption models and the
computation of adsorption energies for key intermediates involved in the associative NRR
pathway, namely *N$_2$, *N$_2$H, and *NH$_3$. Unlike most studies that examine only a
limited number of adsorption sites per material, our work systematically evaluates all accessible
surface sites of the considered IMCs, providing a more comprehensive picture of their
catalytic potential for NRR. We then analyze how changes in IMC composition influence
adsorption configurations and energies. While systematic screening of IMCs for NRR remains
underexplored and existing studies often focus on noble-metal-based systems, here we
investigate materials composed of Co, Ni, Al, Zn, V, Fe, Cu, and Pt, encompassing both
earth-abundant and noble elements. Al and Zn are introduced not for their intrinsic NRR
activity but to probe whether their low affinity can facilitate the desorption of ammonia
when alloyed with transition metals. We further characterize perturbations in linear scaling
relationships (LSRs) across these systems. Subsequently, we employ multiple machine learning
models to predict adsorption energies across adsorption sites by considering different sizes of
the local neighboring shell and using a descriptor set that combines electronic features
(such as s-, p-, and d-band centers and atomic charges) with intrinsic atomic properties.
Notably, we show that by selecting physically meaningful descriptors---particularly those
capturing the local atomic environment and electronic structure---accurate predictions of
adsorption energies can be achieved using a compact set of only 20 key features for *N$_2$H
and *NH$_3$ and 38 features for *N$_2$, enabling the
use of simple and computationally efficient machine learning models. Furthermore, our analysis
reveals that s- and p-orbital characteristics play a significant role in adsorbate-surface
interactions, extending beyond their traditionally recognized importance in systems containing
p-block elements or transition metals with filled or nearly filled d-orbitals. Instead, these
orbitals are shown to actively contribute even to transition metal systems with partially
filled d states, highlighting their broader role in governing adsorption behavior. The
analysis that led to these discoveries is presented in the remainder of this article organized
as follows: Section~2 describes the methods and computational details, Section~3 presents the
results and discussion, and Section~4 summarizes the conclusions.

\section{Methods and Computational Details}

\subsection{DFT Calculations}

DFT calculations were performed using the Vienna Ab initio Simulation Package
(VASP)~\cite{Kresse1996}. The exchange--correlation energy was described using the
Perdew--Burke--Ernzerhof (PBE) functional within the generalized gradient approximation
(GGA)~\cite{Perdew1992,Burke1998}. The projector augmented-wave (PAW)
method~\cite{Blochl1994} was employed with a plane-wave cutoff energy of 500~eV. Spin
polarization was explicitly included for all compounds containing magnetic elements by
setting ISPIN~$=2$. Initial magnetic moments of 6, 4.5, and 3~$\mu_B$ were assigned to Fe,
Co, and Ni, respectively. These values served as initial conditions, while the spin density
and total magnetization were allowed to evolve self-consistently during electronic and
structural relaxation. The initialization values were selected after testing different
initial magnetic moments to reduce the likelihood of convergence to low-spin or
nonmagnetic solutions. The same element-specific initial magnetic moments were applied to
the corresponding atoms in the surface-slab calculations. As a computational consistency
check, the converged bulk magnetizations were compared with the corresponding Materials
Project or literature values, reported in Table~S1. General agreement was obtained, with
Zn$_9$Fe$_4$ being the exception. Because the Materials Project values are also derived
computationally, they are provided for comparison rather than as experimental validation.
Zn$_9$Fe$_4$ is a structurally complex $\gamma$-brass phase containing multiple
symmetry-distinct sites and reported Fe/Zn mixing across different sublattices. A DFT
calculation normally represents this using one specific ordered arrangement of Fe and Zn. A
different arrangement could give each Fe atom different neighbours and therefore produce a
different total magnetic moment. This may explain the greater discrepancy between our
result and the Materials Project value for this compound. For both bulk and surface
optimizations, the convergence thresholds were set to $10^{-5}$~eV for electronic steps and
0.05~eV/\AA{} for ionic relaxations. Surface slabs consisted of 4 atomic layers, with the
bottom 2 layers fixed at their positions and the remaining layers fully relaxed, separated
by a vacuum spacing of 15~\AA{} along the surface normal to suppress interactions between
periodic slab images. A $1\times1\times1$ Monkhorst--Pack k-point grid was used for bulk and
surface structure optimizations. Bulk calculations were performed on a real-space-expanded
supercell prior to applying this mesh: smaller unit cells were expanded by a larger factor,
while larger unit cells, whose Brillouin zones are already small, were used with little or
no further expansion (see Table~S1). A finer $3\times3\times3$ k-point grid was used for the
density-of-states (DOS) calculations. Charge analysis was performed using Bader's quantum
theory of atoms in molecules~\cite{Tang2009,Henkelman2006}.

\subsection{Dataset Generation}

\begin{figure}[htbp]
  \centering
  \includegraphics[width=\linewidth]{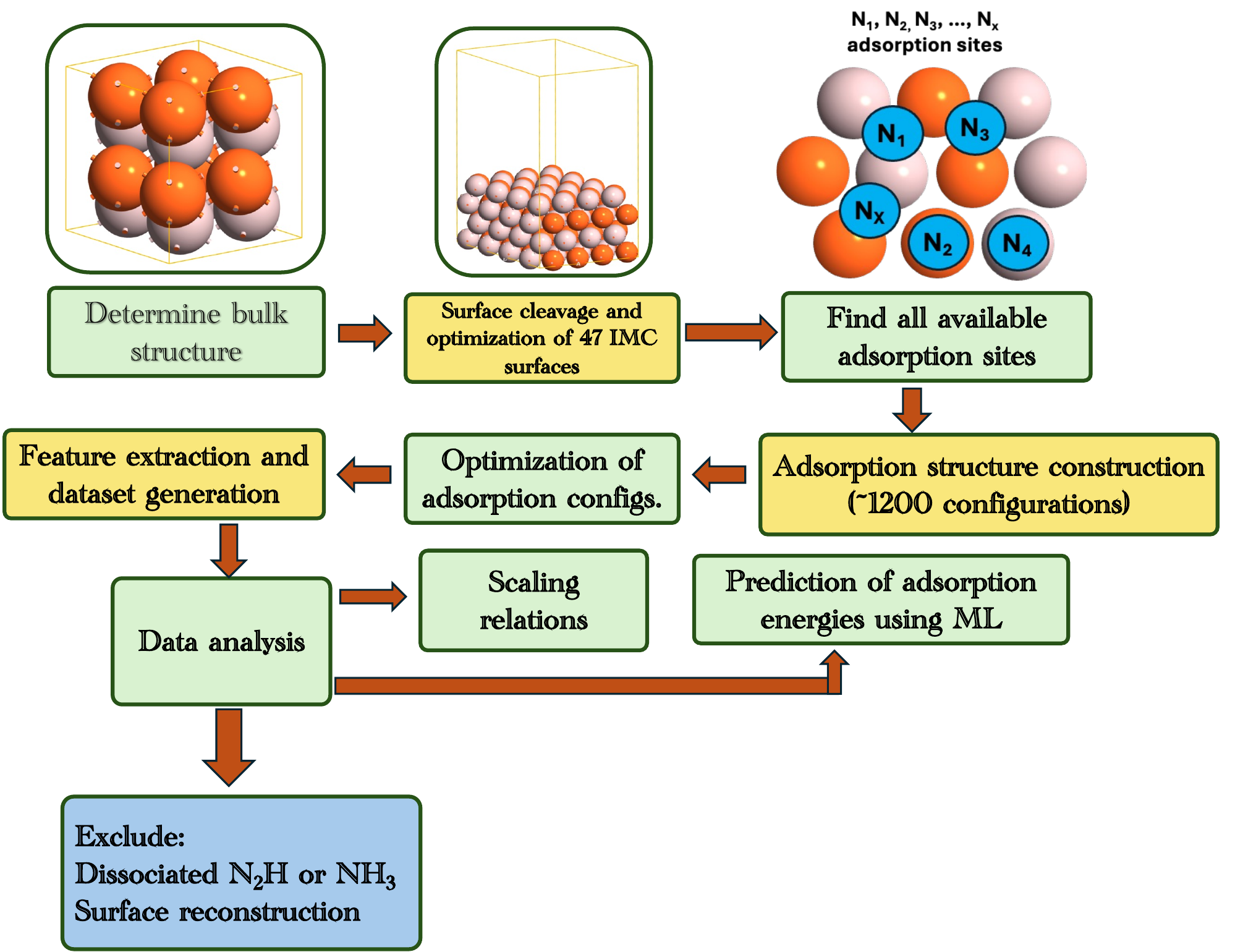}
  \caption{Schematic workflow for high-throughput DFT and machine learning analysis of IMCs.
  The process involves selecting the bulk structures, surface cleavage and optimization,
  identification of all available adsorption sites, and generation of $\sim$1200 adsorption
  configurations. After geometry optimization, a tabular dataset is generated by extracting
  electronic, structural, and intrinsic properties. The scaling relation analysis and machine
  learning prediction of adsorption energies are then performed on the prepared dataset.
  Unstable configurations, such as dissociated N$_2$H or NH$_3$ and surfaces undergoing
  reconstruction, are excluded from the dataset.}
  \label{fig:fig1}
\end{figure}

We selected 47 bimetallic IMCs composed of eight elements (Co, Ni, Al, Zn, V, Fe, Cu, and
Pt), which together cover a wide range of N$_2$ adsorption strengths. These elements were
chosen to span diverse d-electron counts (ranging from $3d^3$ in V to $3d^{10}$ in Cu/Zn
and $5d^9$ in Pt) and electronegativities. Furthermore, Al and Zn were incorporated to
probe whether their lower affinity can facilitate NH$_3$ desorption and how s- and
p-orbital hybridization modulates intermediate binding strength when alloyed with
earth-abundant transition metals (Fe, Co, Ni, V, Cu). The corresponding elemental
properties relevant to catalytic activity are summarized in Table~S2. Table~S1 lists the
Materials Project IDs, crystal system, space groups, and formation energies of the 47
investigated IMC bulk structures obtained from the Materials Project
database~\cite{Horton2025}. For each composition, bulk unit cells were selected based on
their ground-state thermodynamic stability (lowest energy on the convex hull) and verified
against experimentally synthesized ordered phases in the Inorganic Crystal Structure
Database (ICSD). Although only one compound, VCo$_3$, was theoretically predicted to have
a positive formation energy in the list, the Materials Project database indicates that it
has been experimentally synthesized. It is important to note that formation energies
derived from bulk materials cannot be used as direct guidelines for nanocatalyst synthesis.
In practice, ordered nanocatalysts, including many with positive formation energies, can be
obtained through thermal annealing~\cite{Zhou2023ACS,Li2019}. To apply a consistent and
automatable facet-selection procedure across the 47 intermetallic compounds, one
representative crystallographic orientation was selected for each compound based on the
highest-intensity reflection in the simulated XRD pattern generated using VESTA. This
criterion provided a reproducible means of selecting crystallographically characteristic
and experimentally recognizable planes across the chemically and structurally diverse
dataset. For the common intermetallic crystal systems in our dataset (e.g., L1$_0$,
L1$_2$, B2), the highest-intensity XRD reflections coincide directly with low-index surface
facets such as $\{111\}$, $\{110\}$, and $\{100\}$. Specifically, 43 out of the 47
investigated materials correspond directly to these fundamental low-index families. This
includes both direct low-index cuts (e.g., $(1,1,1)$, $(1,1,0)$) and their higher-order
Bragg reflection representations in diffraction software (e.g., $(0,0,2)$ or $(0,2,2)$).
For the remaining 4 materials, selecting the highest XRD intensity facet allows us to
stick strictly to the most experimentally synthesizable and thermodynamically abundant
plane for those specific complex crystal structures. Therefore, this selection criterion
acts as a practical strategy that ties our surfaces to experimentally recognizable crystal
planes while, in the majority of cases, coinciding with the low-index facets. The selected
planes are also listed in Table~S1. The entire calculation workflow is shown in
Fig.~\ref{fig:fig1}. We first used the open-source pymatgen software~\cite{Ong2013} to
identify all the unique adsorption sites and construct adsorption structures on IMC
surfaces. Three adsorbates associated with the associative NRR pathway (*N$_2$, *N$_2$H,
and *NH$_3$) were considered. In total, approximately 1,200 adsorption structures were
included in our high-throughput calculations. The adsorption energies of *N$_2$, *N$_2$H,
and *NH$_3$ molecules on different adsorption sites of IMC surfaces are calculated using
Eq.~\ref{eq:eads}. The N$_2$ and N$_2$H molecules were initially placed horizontally on the
surface, with both nitrogen atoms directly in contact with the metallic surface, and then
optimized using DFT.
\begin{equation}
  E_{\mathrm{ads}} = E_{\mathrm{surface+adsorbate}} - \left(E_{\mathrm{surface}}
  + E_{\mathrm{adsorbate}}\right)
  \label{eq:eads}
\end{equation}

where $E_{\mathrm{ads}}$ is the adsorption energy, $E_{\mathrm{surface+molecule}}$ is the
total energy of the surface with the adsorbate, $E_{\mathrm{surface}}$ is the total energy of
the clean surface, and $E_{\mathrm{adsorbate}}$ is the energy of the isolated molecule.

\FloatBarrier

\subsection{Interpretable Machine Learning Approach for Adsorption Energy Prediction}

Recent advances in machine learning have underscored the importance of feature selection in
developing accurate and interpretable predictive models for catalysis. In heterogeneous
catalysis, the adsorption site dictates the reaction pathway, and therefore the chosen
descriptors must capture both the local structural and electronic environment of the site
while providing chemically meaningful insights. With this motivation, we constructed a
dataset that incorporates electronic, structural and intrinsic material descriptors of the
adsorption environment. Specifically, we calculated the total d-band center of the whole
structure ($\varepsilon_{d,\mathrm{total}}$) as well as the d-band centers of atoms up to
the 21st nearest neighbor around each adsorbate ($\varepsilon_{d,\mathrm{local}}$) to
understand the effect of the neighboring shell size on the material's adsorption
properties. We also considered the s- and p-band characteristics of atoms~\cite{Chen2022}.
Another property derived from the projected density of states (PDOS) is the orbital band
filling, defined as the integrated weight of the orbital DOS over the selected energy
window ($-10$ to $10$~eV). The equations for the s-, p-, and d-band centers and band
filling are provided in Eqs.~\ref{eq:bandcenter} and~\ref{eq:bandfilling}, respectively.
\begin{equation}
  \varepsilon_{\alpha} = \frac{\displaystyle\int_{-10}^{+10} n_{\alpha}(\epsilon)\,\epsilon
  \,d\epsilon}{\displaystyle\int_{-10}^{+10} n_{\alpha}(\epsilon)\,d\epsilon}
  \label{eq:bandcenter}
\end{equation}

where $\varepsilon_{\alpha}$ is the band center of orbital $\alpha$, and
$n_{\alpha}(\epsilon)$ denotes the PDOS of orbital $\alpha$ as a function of energy
$\epsilon$ (in eV). The numerator represents the total energy-weighted PDOS, and the
denominator normalizes it, yielding the average energy position of the orbital states within
the selected window.
\begin{equation}
  f_{\alpha} = \int_{-10}^{+10} n_{\alpha}(\epsilon)\,d\epsilon
  \label{eq:bandfilling}
\end{equation}

where $f_{\alpha}$ is the orbital band filling. Band filling describes the total electronic
population of an orbital (s-, p-, d-) within the specified energy range. Bader charges ($q$)
of neighboring atoms of adsorbates are also included as electronic features, obtained from DFT
calculations. To describe the structural and intrinsic properties of adsorption sites, we
included atomic number ($Z$), atomic radius ($r$), electron affinity (EA), electronegativity
($\chi$), covalent radius ($r_{\mathrm{cov}}$), atomic density ($\rho$), first ionization
energy (IE), and atomic mass ($M$), of up to 21 neighboring atoms, along with their
distances ($d$) to the adsorbate. For the first nearest neighbor atom to adsorbate ($n_0$),
these properties are included as individual features. For the rest of the surrounding
neighbor shell ($n_1$ through $n_{20}$), rather than listing each neighbor
individually---which would be order-sensitive and grow in dimensionality with the number of
atom shells---we computed six aggregated statistics across the shell for each property: the
arithmetic mean ($\mu$), standard deviation ($\sigma$), minimum, maximum, range ($\Delta$),
and distance-weighted mean ($\mu_w$) (closer atoms contribute more strongly) according to the
following equation:
\begin{equation}
  \mu_w = \frac{\sum_i (v_i / d_i)}{\sum_i (1/d_i)}
  \label{eq:distweight}
\end{equation}

where $v_i$ is the property value of neighbor $i$ and $d_i$ is its distance to the binding
site. This results in a constant feature dimensionality of 72 regardless of the number of
neighbor shells considered, making the feature set order-invariant and avoiding the rapid
growth in feature dimensionality that occurs when each neighbor atom is represented
individually. In total, the feature set comprises two global structural descriptors (crystal
structure type and symmetry group), ten $n_0$ individual descriptors, 48 shell statistics
(8~properties $\times$ 6~statistics), and 12~DOS descriptors (please see Table~S3 for a full
description of all descriptors used). Prior to training, extreme outliers in adsorption
energy---which are identified as samples beyond three standard deviations from the mean
($|z|>3$) of the target distribution---were removed from the dataset for each adsorbate. To
assess the sensitivity of model performance to this step, we additionally trained and
evaluated the full ML pipeline on the unfiltered dataset (i.e., without removing $|z|>3$
samples) for each adsorbate. After outlier removal, the training/test datasets comprised
251, 288, and 365 samples for *N$_2$, *N$_2$H, and *NH$_3$, respectively, compared to 255,
292, and 375 samples in the corresponding unfiltered datasets (4, 4, and 10 samples excluded
per adsorbate).

A diverse set of machine learning algorithms was employed to capture the relationship between
the feature set and adsorption energy. Linear models included Ridge Regression (Ridge), Lasso
Regression (Lasso), Elastic Net (ENet), and Huber Regression (Huber). Kernel- and
distance-based models included Support Vector Regression with an RBF kernel (SVR), Kernel
Ridge Regression (KRR), and k-Nearest Neighbors (KNN). Tree-based methods comprised Decision
Tree (DT), Random Forest (RF), Gradient Boosting (GB), and Extra Trees (ET). A Multi-Layer
Perceptron neural network (MLP) with two hidden layers of 128 and 64 neurons was also
included. The advanced gradient boosting implementation XGBoost (XGB) was also evaluated. The
dataset was split into 80\% training and 20\% test sets using a fixed random seed to ensure
reproducibility. The test set was treated as a sealed holdout and evaluated after all model
selection decisions were finalized. For models sensitive to feature scaling,
including Ridge, Lasso, ENet, Huber, SVR, KRR, MLP, and KNN, the features were standardized
using the StandardScaler module implemented in scikit-learn library to account for the wide
range in the feature numerical scales~\cite{Buitinck2013}:
\begin{equation}
  x' = \frac{x - \mu}{\sigma}
  \label{eq:normalization}
\end{equation}
where $x$ is the feature value, $\sigma$ is the standard deviation of the feature computed on
the training set, and $\mu$ is the feature mean. The DT, RF, GB, ET, and XGB models were used
without scaling, as tree-based algorithms are not sensitive to the scale of the variables.
The hyperparameter setting used for each algorithm are listed in Table~S4.

Given the limited training set size relative to the 72 engineered features, dimensionality
reduction was applied to avoid overfitting. The full preprocessing pipeline consisted of
four sequential steps fitted exclusively on the training portion of each cross-validation
fold: (1)~VarianceThreshold (threshold~$= 0.01$) to remove near-constant features,
(2)~DropHighCorr (threshold~$= 0.95$) to remove one feature from each highly correlated
pair based on Pearson correlation, and (3)~SelectKBest with the F-regression scoring
function, which ranks all remaining features by their univariate linear correlation with
the adsorption energy and retains up to the top $K$ (feature-selection budget) of them
before any model sees the data. To determine an appropriate feature-selection budget, this
pipeline was repeated while varying $K$ ($K = 10, 20, 30, 40, 50, 60$, and $72$) and
evaluating cross-validation performance at each value for all three adsorbates. Because
VarianceThreshold and DropHighCorr already remove a substantial fraction of the 72
engineered features before SelectKBest is applied, requesting $K = 72$ does not restore the
full feature set for any adsorbate or outlier condition: at most 38 features survive for
*N$_2$, 35 for *N$_2$H, and 37 for *NH$_3$ under the outlier-removed ($|z| \leq 3$)
condition, and at most 40 for *N$_2$, 39 for *N$_2$H, and 37 for *NH$_3$ under the
unfiltered condition. Once $K$ reaches these adsorbate- and condition-specific ceilings,
SelectKBest simply retains every surviving feature rather than narrowing them down further,
so larger budgets do not change the outcome. As a robustness check, this $K$-sweep was
additionally repeated using recursive feature elimination (RFE) with a ridge-regression
base estimator in place of the F-regression scoring function. Once $K$ reached the ceilings
described above, RFE and SelectKBest converged to numerically identical feature sets and
cross-validation performance for every adsorbate and outlier condition; below these
ceilings, cross-validation $R^2$ differed by no more than 0.04 between the two methods
across all three adsorbates and both outlier conditions (Table~S5). Step~(4) applied
StandardScaler for models sensitive to feature scale. All four steps were encapsulated as a
single scikit-learn Pipeline object together with the model, so that validation samples were
never used to fit any preprocessing step and they were only passed through the already-fitted
transformations before prediction, ensuring that the cross-validation $R^2$ reflects
predictive performance on held-out adsorption sites within the investigated material space.

Model selection was performed by sweeping over the number of neighbor shells $k$ from 0 to
20, training and evaluating all models at each level. Performance was assessed using Repeated
K-Fold cross-validation with 5~splits and 10~repeats (50~total evaluations per model per
neighbor level) on the training set only, yielding stable estimates of predictive
performance across held-out adsorption sites. The
best model and neighbor level were selected based on cross-validation $R^2$, and the
corresponding model was retrained on the full training set before final evaluation on the
sealed test set. The performance of each model was evaluated using the mean absolute error
(MAE), root mean squared error (RMSE), and coefficient of determination ($R^2$), defined as:
\begin{equation}
  \mathrm{MAE} = \frac{1}{N}\sum_{i=1}^{N}\left|y_i - \hat{y}_i\right|
  \label{eq:mae}
\end{equation}

\begin{equation}
  \mathrm{RMSE} = \sqrt{\frac{1}{N}\sum_{i=1}^{N}\left(y_i - \hat{y}_i\right)^2}
  \label{eq:rmse}
\end{equation}

\begin{equation}
  R^2 = 1 - \frac{\displaystyle\sum_{i=1}^{N}\left(y_i - \hat{y}_i\right)^2}
  {\displaystyle\sum_{i=1}^{N}\left(y_i - \bar{y}\right)^2}
  \label{eq:r2}
\end{equation}

where $y_i$ is the adsorption energy obtained from DFT calculations, $\hat{y}_i$ is the
predicted value, $\bar{y}$ is the mean of the DFT values, and $N$ is the number of data
points. The performance of the machine learning model is considered better when the RMSE and
MAE values are close to zero, while the $R^2$ value is close to 1. The influence of
individual features on predicted adsorption energies was evaluated using SHapley Additive
exPlanations (SHAP)~\cite{Lundberg2017}. For tree-based models, exact SHAP values were
computed using TreeExplainer; for linear models, LinearExplainer was used; and for kernel and
distance-based models, KernelExplainer with 50 background samples drawn from the training set
was applied.

\section{Results and Discussions}

\subsection{Adsorption Energy Analysis}

\begin{figure}[htbp]
  \centering
  \includegraphics[width=\linewidth]{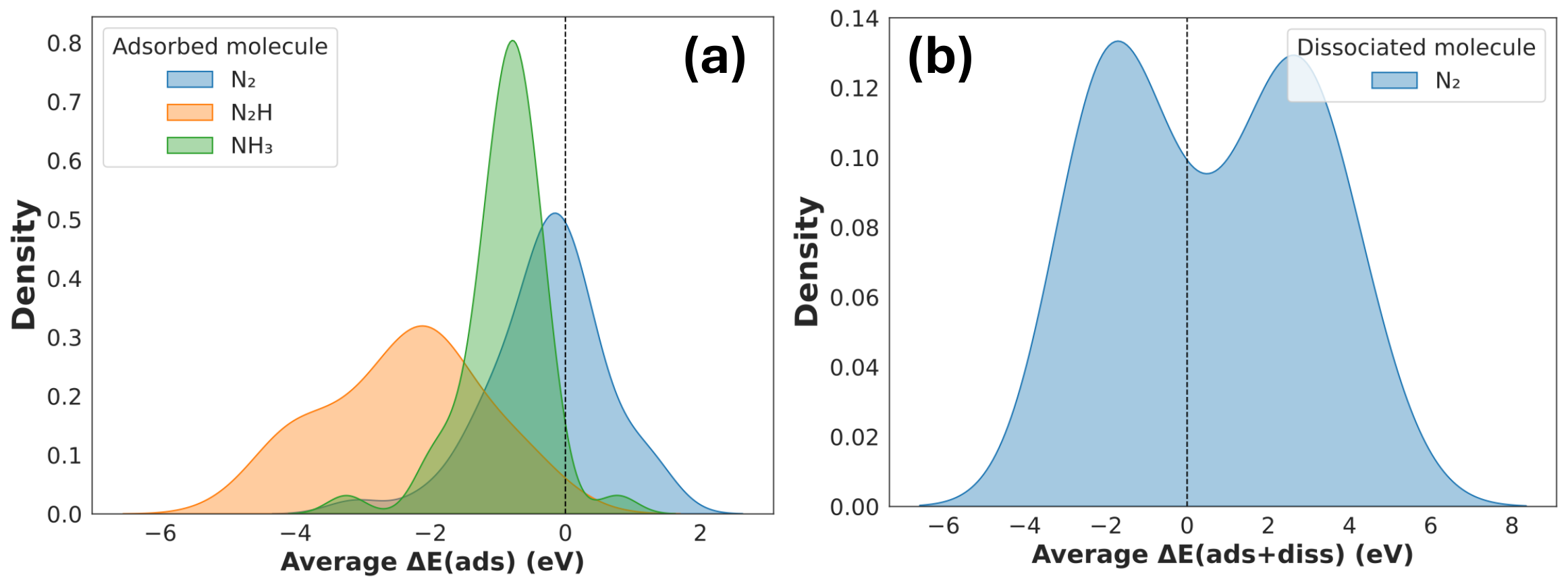}
  \caption{Distribution of average bonding energies for (a) adsorbed molecules and
  (b) dissociated N$_2$ across IMCs.}
  \label{fig:fig2}
\end{figure}

Fig.~\ref{fig:fig2} shows the distribution of the average adsorption and dissociation
energies of key NRR intermediates across all adsorption sites of the investigated intermetallic
compounds (IMCs). In Fig.~\ref{fig:fig2}(a), the average adsorption energies of N$_2$,
NH$_3$, and N$_2$H are plotted, where each data point represents the mean value taken over
all adsorption sites of a given alloy. The resulting distributions highlight clear differences
between adsorbates: while NH$_3$ adsorption energies cluster closer to zero, indicating
generally weaker binding, N$_2$H has a broader distribution shifted toward more negative
values, suggesting that many alloys stabilize this intermediate more strongly on average. The
adsorption energies of N$_2$ are more widely dispersed than those of NH$_3$, but show a
narrower range compared to N$_2$H. In Fig.~\ref{fig:fig2}(b), the distribution of N$_2$
dissociative adsorption energies is shown, indicating that N$_2$ dissociation energy has a
wide distribution across alloys. This is due to the fact that nitrogen adsorption is sensitive
to its local environment and small changes in its neighboring atoms can cause significant shifts
in bonding energies of N$_2$, as shown in our earlier work~\cite{Agharezaei2025}.

\FloatBarrier

Fig.~S1 shows the distribution of adsorption energies for key molecular and atomic adsorbates
across the IMCs. Tables~S6--S8 summarize the statistical analysis of adsorption properties
across IMCs. Fig.~S1a--c shows the distribution of adsorption energies of N$_2$, N$_2$H, and
NH$_3$ molecules, and Fig.~S1d shows the distribution of N$_2$ dissociative adsorption
energies (*2N). Each plot illustrates the spread of adsorption energies over all adsorption
sites within a given alloy, capturing both the average bonding strength and the variability
in adsorption energies across different sites.

43 out of 47 alloys have at least one site that adsorbs molecular nitrogen and most alloys
have binding energies between $-2.0$ and $0.5$~eV (Fig.~S1a). The dissociation of nitrogen, however, is
less common, occurring in only 21 alloys, and is typically endothermic with a median energy of
$+2.21$~eV. This confirms that direct dissociation is possible only in a limited number of
systems and that the dissociative NRR pathway is less probable overall. A small number of
alloys, such as AlV$_3$ ($\Delta E(\mathrm{ads}) = -3.36$~eV), have exceptionally strong N$_2$
adsorption on the surface. N$_2$H adsorption (Fig.~S1b) is generally stronger, with most
adsorption energies of IMCs falling between $-3.0$ and 0~eV and a median energy of
$\approx -2.34$~eV. This strong stabilization is crucial for lowering the hydrogenation barrier,
although overly negative values ($\leq -2.5$~eV) indicate potential desorption challenges and
surface poisoning~\cite{Qian2020}. NH$_3$ (Fig.~S1c) has a median binding energy of
$\approx -0.81$~eV. While most values lie between $-2.0$ and 0~eV, some alloys bind NH$_3$ too
strongly ($\leq -2.0$~eV), which is unfavorable since efficient NRR requires facile desorption
of the final product. Indeed, recent computational studies suggest that NH$_3$ adsorption
energies in the range of $-1.0$ to $-0.5$~eV are optimal for avoiding surface
poisoning~\cite{Tezak2023}.

The adsorption energy distributions further reveal that N$_2$H exhibits only modest variation
across adsorption sites within the same alloy, while its average value increases steadily across
different IMCs. NH$_3$ similarly displays a narrow site-to-site spread but shows less
significant variation between alloys. In contrast, N$_2$ adsorption energies vary widely even
within a single alloy, underscoring the strong sensitivity of N$_2$ to its local adsorption
geometry and initial positioning on the surface. Importantly, the wide ranges observed for
several IMCs, represented in Fig.~S1, highlight that these materials cannot be represented by
a single adsorption energy, instead, they offer various adsorption environments, some of which
may fall within the ``ideal'' window for NRR activity. For example, while the N$_2$
adsorption is typically relatively weak, several alloys possess rare sites with
$\Delta E(\mathrm{N}_2) \leq -3$~eV. Such outlier sites may dominate the catalytic activity
under reaction conditions, emphasizing that heterogeneity is a feature to be explored (not
avoided). This observation supports the view that multicomponent alloys can host exceptional
adsorption sites that outperform average LSRs~\cite{Xiao2024}. Consequently, adsorption energy
distributions, rather than single-valued descriptors, provide a more reliable framework for
the rational design of efficient NRR catalysts.

\begin{figure}[htbp]
  \centering
  \includegraphics[width=\linewidth]{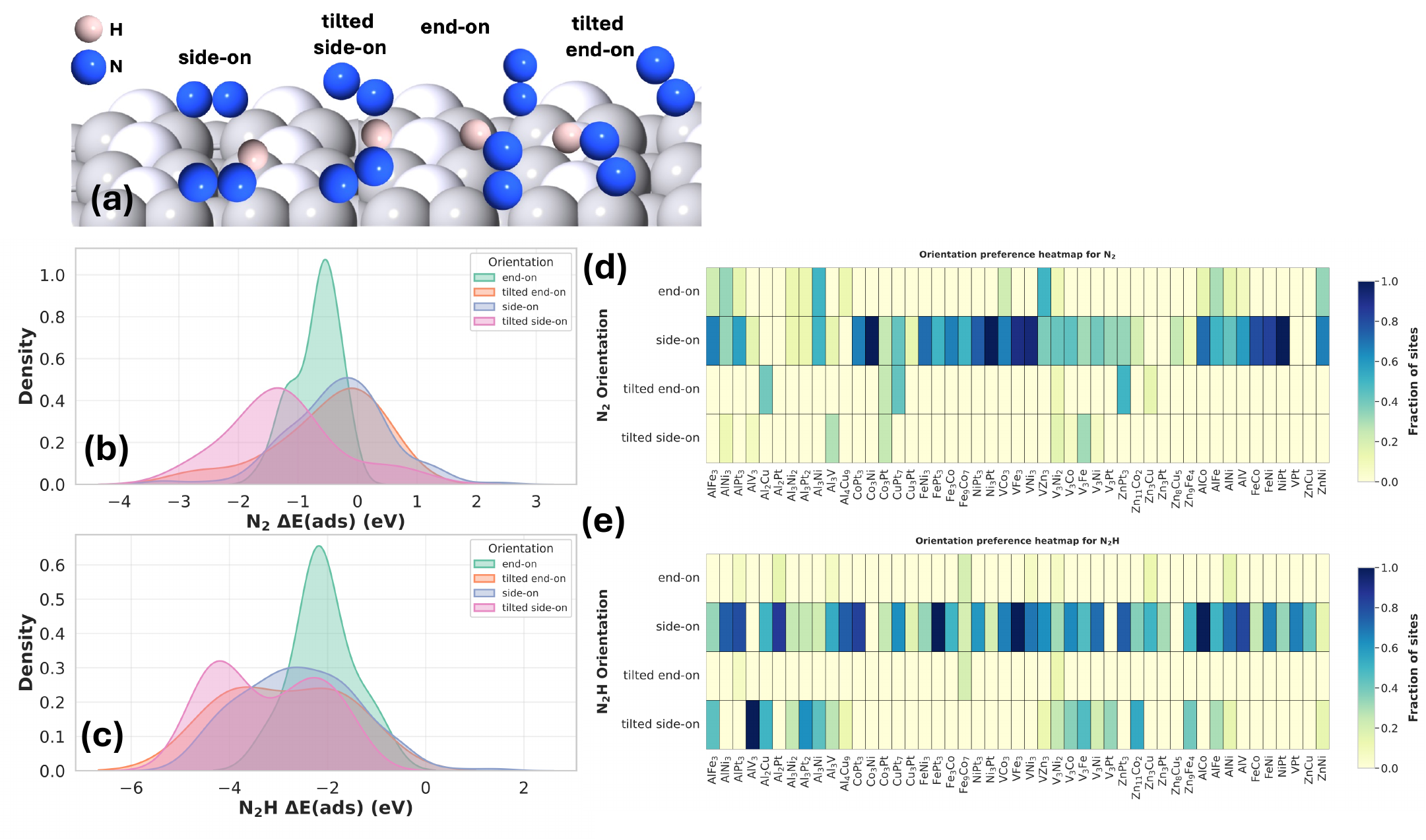}
  \caption{(a) Different orientations of *N$_2$ and *N$_2$H molecules on the surface of IMCs;
  Distribution of adsorption energies for (b) *N$_2$ and (c) *N$_2$H molecules in different
  orientations on IMC surfaces; Heatmaps of adsorption orientation preferences for (d) *N$_2$,
  and (e) *N$_2$H across IMCs. The color scale represents the fraction of adsorption sites in
  each alloy that adsorb the molecule in each orientation. Orientations of the adsorbed
  molecules are classified as end-on, tilted end-on, side-on, and tilted side-on.}
  \label{fig:fig3}
\end{figure}

Fig.~\ref{fig:fig3}(a) shows the different orientations of *N$_2$ and *N$_2$H molecules on the
surfaces of IMCs after DFT optimization. When the angle between the N--N axis and the surface
normal is 90$^\circ$, 0$^\circ$, greater than 45$^\circ$, or less than 45$^\circ$, the
corresponding orientations are defined as side-on, end-on, tilted side-on, and tilted end-on,
respectively. Fig.~\ref{fig:fig3}(b--c) shows the distribution of adsorption energies for
different orientations of *N$_2$ and *N$_2$H molecules on the surfaces of IMCs. For *N$_2$, the
end-on orientation has a narrower and sharper energy distribution, indicating stronger binding
to the surface that is probably less dependent on neighboring atoms. In contrast, side-on
orientations display a broader distribution, reflecting more variability in binding strength
depending on the surface structure. Interestingly, even a slight tilt from the end-on
orientation significantly reduces the stability of the *N$_2$ molecule, with the adsorption
energy increasing by approximately 0.5~eV at the peak of the tilted end-on distribution.
Tilting a side-on *N$_2$ molecule in a way that one nitrogen atom moves slightly farther from
the surface enhances its stability, lowering the adsorption energy by about 1.5~eV at the
tilted side-on peak. These observations suggest that the most stable configuration for *N$_2$
on IMCs is typically a tilted side-on orientation. *N$_2$H also shows a sharp adsorption energy
distribution for the end-on orientation. However, only a small number of sites (just 12 across
all IMCs) adsorb N$_2$H in either the end-on or tilted end-on configurations. In contrast,
side-on adsorption of N$_2$H shows a broader energy distribution, with a significant number of
sites showing a strong binding, with adsorption energies below $-2$~eV. Comparing the plots
for *N$_2$ and *N$_2$H, it is clear that *N$_2$H is easily adsorbed on all the sites, irrespective
of the *N$_2$H orientation, in contrast to *N$_2$. Fig.~\ref{fig:fig3}(d--e) shows the fraction
of adsorption sites that adsorb N$_2$ or N$_2$H molecules in a specific orientation. The
majority of sites adsorb both N$_2$ and N$_2$H in a side-on orientation.

\FloatBarrier

\begin{figure}[htbp]
  \centering
  \includegraphics[width=0.85\linewidth]{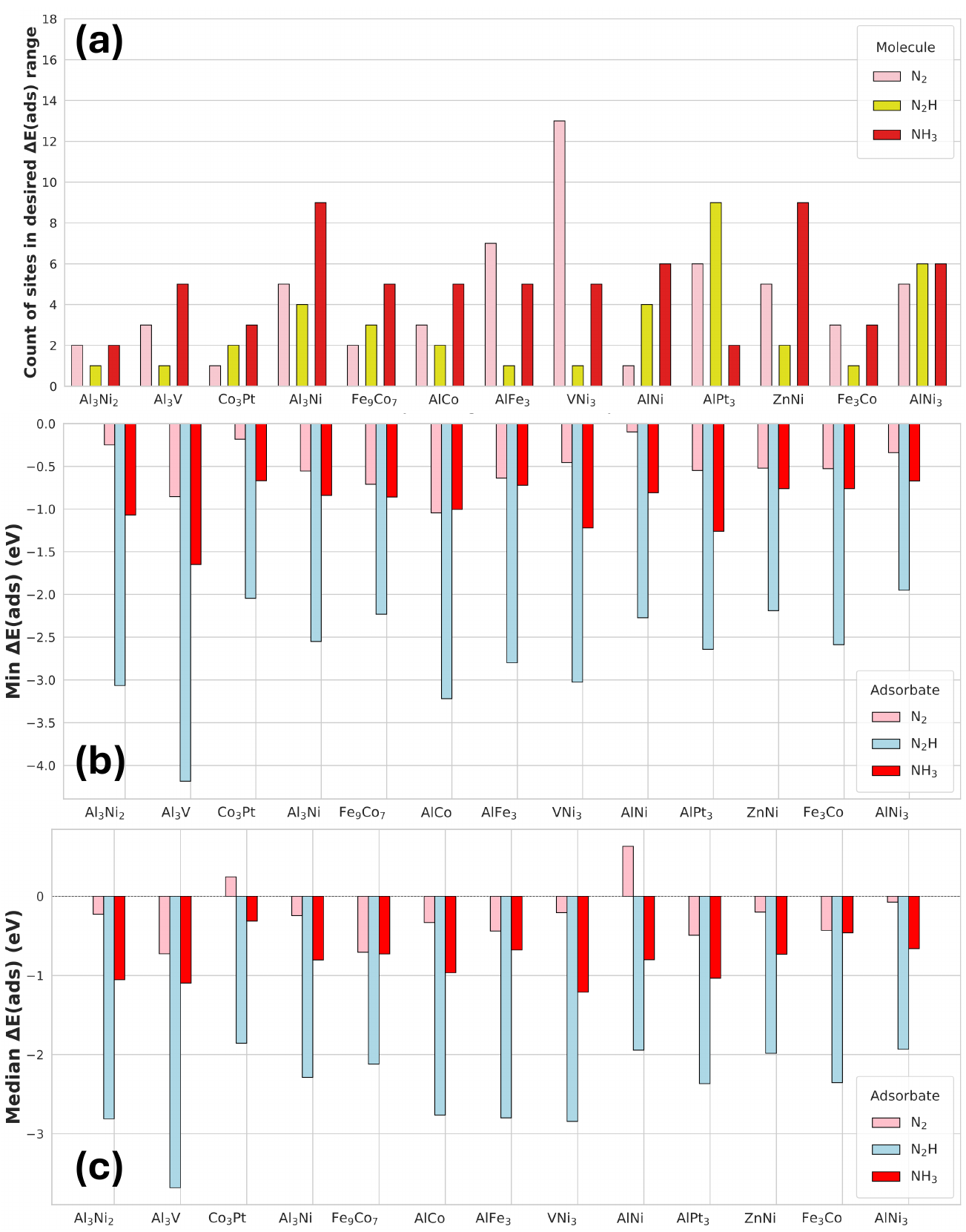}
  \caption{Chosen IMCs that possess at least one adsorption site for each adsorbate falling
  within the specified adsorption energy window. (a) Number of adsorption sites for N$_2$,
  N$_2$H, and NH$_3$ on IMCs with adsorption energies within the defined windows:
  *N$_2$ ($-1.0$ to $0.0$~eV), *N$_2$H ($-2.5$ to $-1.5$~eV), and *NH$_3$ ($-1.0$ to
  $0.0$~eV). (b) Minimum and (c) median adsorption energies for *N$_2$, *N$_2$H, and *NH$_3$
  of the same IMCs.}
  \label{fig:fig4}
\end{figure}

To identify the most effective catalysts among the IMCs, adsorption energy windows were
defined following strategies reported in previous studies. For *N$_2$, moderate adsorption is
desired, with an energy range of $-1.0$ to $0.0$~eV~\cite{Shen2024,Hoskuldsson2023}. For
*N$_2$H, a range of $-2.5$ to $-1.5$~eV was chosen, as adsorption stronger than $-2.5$~eV
may hinder further reaction steps~\cite{Qian2020}. Although no explicit range for NH$_3$
desorption has been reported, an energy window of $-1.0$ to $0.0$~eV was used here, as
NH$_3$ should desorb easily from the surface. Fig.~\ref{fig:fig4} shows the 13 out of 47
investigated IMCs that contain adsorption sites for N$_2$, N$_2$H, NH$_3$ with energies
falling within the optimal range. Fig.~\ref{fig:fig4}(a) shows the number of sites in each
IMC that have adsorption energy values for different adsorbates within these target windows.
Fig.~\ref{fig:fig4}(b--c) display the minimum and median adsorption energies across different
adsorption sites in the same IMCs. Al$_3$Ni$_2$, Co$_3$Pt, Al$_3$Ni, AlCo, VNi$_3$, AlNi,
ZnNi, and AlNi$_3$ exhibit weak *N$_2$ adsorption, indicating limited N$_2$ activation. In
contrast, Al$_3$V binds *N$_2$H too strongly, which may impede its further hydrogenation.
Fe$_9$Co$_7$ and Fe$_3$Co exhibit well-balanced adsorption energies for
all intermediates, making them the most promising IMCs for catalytic NRR activity among the
tested IMCs based on the selected adsorption energy window.

\FloatBarrier

\begin{figure}[htbp]
  \centering
  \includegraphics[width=\linewidth]{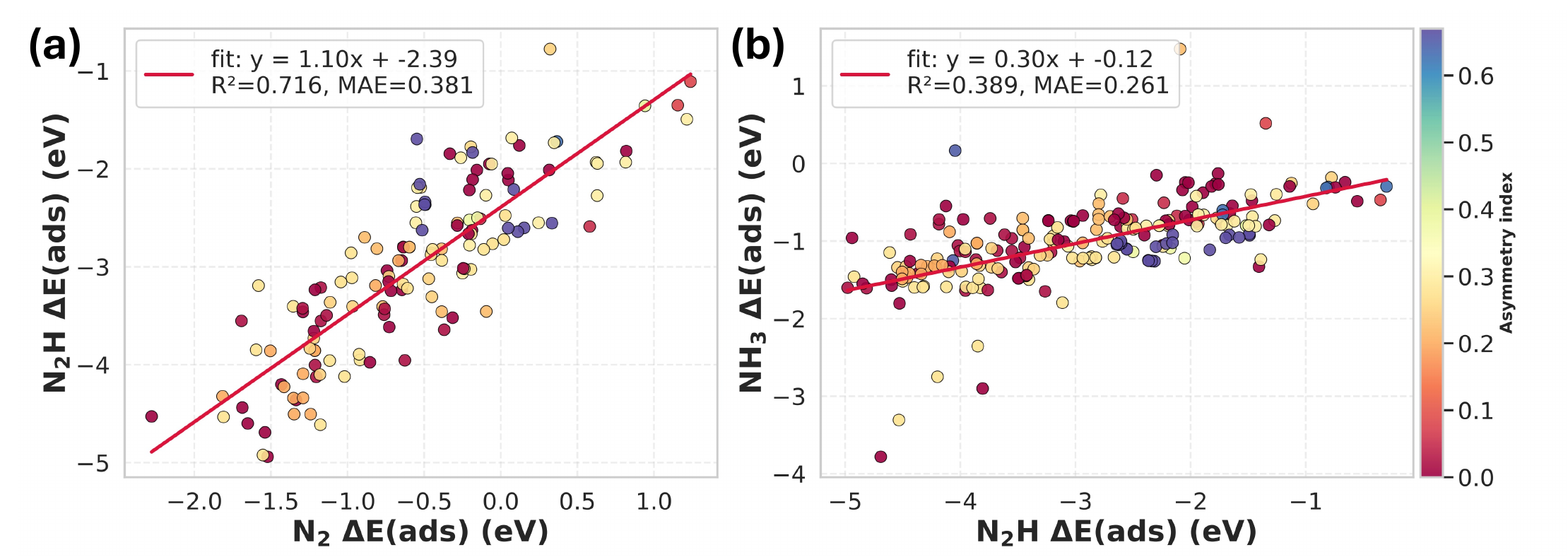}
  \caption{Scaling relationships between adsorption energies of key intermediates on IMCs:
  (a) N$_2$H vs.\ N$_2$, and (b) NH$_3$ vs.\ N$_2$H. Data points are colored according to
  the asymmetry index, calculated as the difference between the highest and the lowest
  electronegativity values among the adsorbate's four nearest neighbors at the
  adsorption site. The fitted red lines highlight the degree of correlation, with $R^2$
  and MAE values reported in each panel.}
  \label{fig:fig5}
\end{figure}

Recent studies of NRR have emphasized the limitations of conventional volcano plots. These
plots typically exhibit strong linear scaling relationships (LSRs) between the adsorption
energies of reaction intermediates~\cite{Exner2022,He2025}. In such cases, the adsorption
energies of intermediates correlate closely with that of N$_2$, which restricts the ability
of the catalyst to simultaneously activate N$_2$ and desorb NH$_3$. Breaking these linear
relations, however, opens new pathways for enhancing catalytic activity by allowing the
adsorption energies of individual intermediates to be tuned independently.
Fig.~\ref{fig:fig5} illustrates the LSRs between the adsorption energies of N$_2$, N$_2$H,
and NH$_3$ on the surfaces of IMCs. Each data point corresponds to the adsorption energies of
two intermediates on the same adsorption site of a given IMC---for example, site 1 for both
N$_2$ and N$_2$H is paired along the $x$- and $y$-axes, respectively. The total number of
data points in Fig.~\ref{fig:fig5} is smaller than the total number of computed adsorption
sites ($\sim$1200), as only configurations in which both species could be stably adsorbed on
the same site were considered. Sites where adsorption was unstable or led to dissociation of
either adsorbate were excluded to ensure a consistent and physically meaningful comparison
across intermediates. Each point is colored by the asymmetry index, calculated as the
difference between the maximum and minimum electronegativity values of the adsorbate's four
nearest neighbors at the adsorption site. The adsorption energies of *N$_2$ and *N$_2$H remain
highly correlated, as stronger *N$_2$ binding leads to stronger *N$_2$H binding as well
(Fig.~\ref{fig:fig5}(a)). Generally, sites with the lowest asymmetry index (red dots) lie
closer to the fitted line, indicating that the adsorption energies of N$_2$ and N$_2$H are
more linearly correlated on sites where the neighboring atoms are of the same element. In
contrast, sites with higher asymmetry index values deviate further from the line. This
observation is consistent with the reported literature, which shows that multielement surface
sites do not fully align with conventional volcano trends. Importantly, sites with higher
asymmetry index values are also closer to the optimum adsorption energies for *N$_2$
($-1.0$ to $0.0$~eV) and *N$_2$H ($-2.5$ to $-1.5$~eV)~\cite{Qian2020,Shen2024,Hoskuldsson2023}.

Fig.~\ref{fig:fig5}(b) shows the correlation between NH$_3$ and N$_2$H adsorption energies.
There is no strong correlation between these intermediates (including between NH$_3$ and
N$_2$, since N$_2$ is correlated with N$_2$H, but N$_2$H is not correlated with NH$_3$). This
decoupling opens opportunities for designing catalysts that maintain strong N$_2$ or N$_2$H
binding while still enabling easy NH$_3$ desorption due to weaker binding.

\FloatBarrier

\begin{figure}[htbp]
  \centering
  \includegraphics[width=\linewidth]{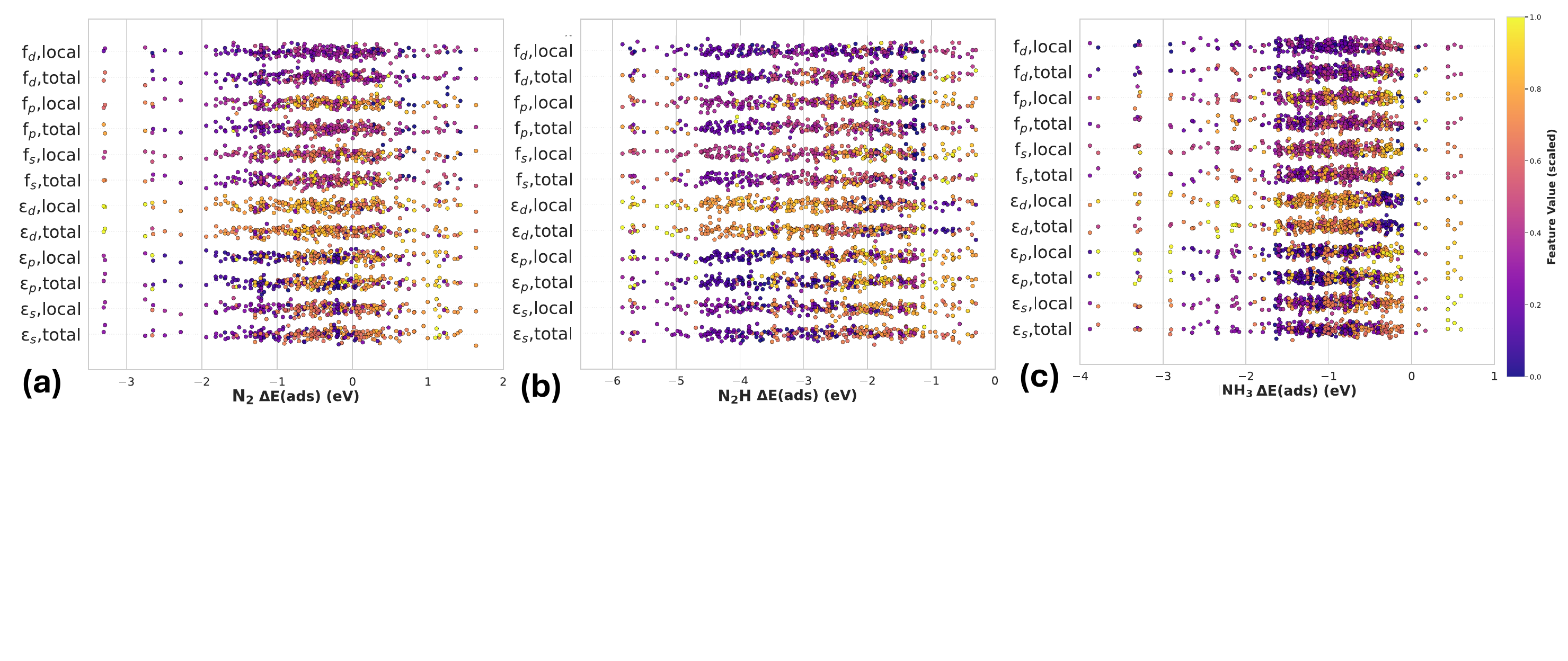}
  \caption{Electronic feature-wise adsorption energy distributions for (a) *N$_2$, (b) *N$_2$H,
  and (c) *NH$_3$ on all IMCs; each horizontal row corresponds to one descriptor
  (PDOS-derived parameter). Points represent individual adsorption sites, plotted by their
  adsorption energy ($x$-axis) with vertical jitter to show density. Point color shows the
  scaled feature value (min--max normalized), revealing how each electronic descriptor
  varies across the adsorption-energy range for each molecule. Local properties refer to
  features computed based on the first four nearest neighbors of the adsorbate, such as local
  d-band filling, or local d-band center.}
  \label{fig:fig6}
\end{figure}

To further understand the role of electronic descriptors in governing adsorption energies,
Fig.~\ref{fig:fig6} presents the distributions of adsorption energies as a function of
individual electronic features. In this plot, orbital-related descriptors are expressed using
compact notation: $\varepsilon_{b,\mathrm{local}}$ denotes the local band center of orbital
$b$ (s, p, or d), calculated from the projected density of states (PDOS) of the
nearest-neighbor atoms of the adsorbate, while $\varepsilon_{b}$ is the total band center for
the entire system. Likewise, $f_{b,\mathrm{local}}$ and $f_{b,\mathrm{total}}$ indicate local
and global orbital fillings, respectively. To ensure consistency across descriptors, local
properties were computed using only the four nearest neighbors to the adsorbate.

Consistent with the trends observed in Fig.~\ref{fig:fig2}, which shows the distribution of
average adsorption energies across all sites in different IMCs, Fig.~\ref{fig:fig6} shows that
N$_2$ and N$_2$H have broader distributions of adsorption energies, whereas NH$_3$ is more
narrowly clustered. Overall, these plots do not suggest a clear distinction between local and
total band descriptors in their relationship to adsorption energy. A more rigorous evaluation
of these effects is provided later through SHAP-based interpretability analysis of machine
learning models. In Fig.~\ref{fig:fig6}(a), for N$_2$ adsorption, there is no obvious direct
correlation between d-band properties (both center and filling) and adsorption energies as
similar d-band values can correspond to a wide range of adsorption strengths. In contrast,
lower values of the local p-band filling ($f_{p,\mathrm{local}}$) appear to correlate with
stronger adsorption (i.e., more negative energy), a pattern that is not as strong for the
total p-band filling. Similar trends are observed for the s- and p-band centers, suggesting
that s and p-band centers may play a more significant role than the d-band in governing
N$_2$ adsorption in IMCs. The s-band and p-band centers also play a significant role in the
adsorption energies of N$_2$H (Fig.~\ref{fig:fig6}(b)). For NH$_3$ (Fig.~\ref{fig:fig6}(c)),
band fillings do not show a strong relationship with adsorption energy, and the impact of the
s-, p-, and d-band centers is less clear compared to N$_2$ and N$_2$H. In this case,
adsorption energies are more heavily clustered, and different electronic features correspond
to a narrow range of adsorption energy values, making it difficult to draw firm conclusions
from this plot alone. Overall, N$_2$ and N$_2$H adsorption energies show broader distributions
than NH$_3$, with s- and p-band centers influencing adsorption more strongly than d-band
properties.

\FloatBarrier

\subsection{Interpretable Machine Learning}

As shown in Fig.~\ref{fig:fig5}, there is no correlation between NH$_3$ adsorption energy,
whose optimum value is critical to catalyst activity, with other intermediates, thus
invalidating the use of *N$_2$ or *N$_x$H$_x$ as descriptors of the reaction. Therefore, we
developed ML models to better understand the catalyst activity toward different
intermediates. This approach helps to identify the key features in the selected materials and
provides a basis for developing future models for screening additional IMCs. The set of
machine learning algorithms was trained using features constructed for different neighbor
shells with $k$ ranging from 0 to 20 (corresponding to the first to the 21st nearest
neighboring atoms around the adsorbate) to predict the adsorption energies of *N$_2$, *N$_2$H,
and *NH$_3$. For each value of $k$, the feature set comprised two global structural
descriptors, ten individual descriptors of the binding site atom, six aggregated statistics
of each intrinsic property across the neighbor shell, and twelve DOS descriptors, yielding
72 features in total. For example, at $k = 3$, the feature set includes the intrinsic atomic
properties of the first nearest neighbor atom to the adsorbate ($n_0$), such as
electronegativity, electron affinity, covalent radius, first ionization energy, and Bader
charge, which are treated as separate descriptors. In addition, six aggregated statistical
features (mean, standard deviation, minimum, maximum, range, and distance-weighted mean) are
computed for each intrinsic atomic property across the surrounding neighbor shell ($n_1$ to
$n_3$). Electronic descriptors derived from the PDOS are also included, comprising the total
and local s-, p-, and d-band centers and fillings evaluated up to the fourth neighbor shell.

\begin{figure}[htbp]
  \centering
  \includegraphics[width=\linewidth]{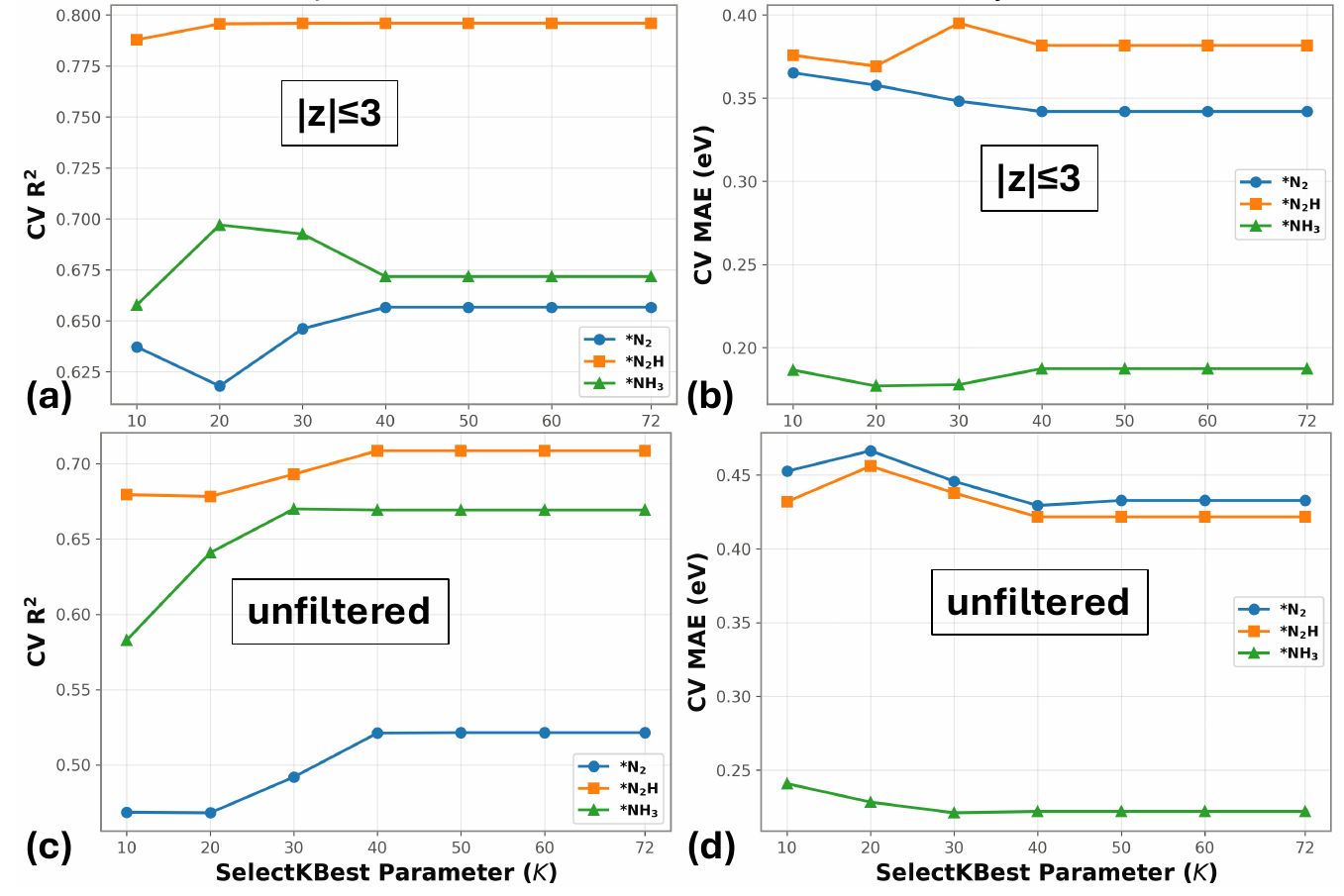}
  \caption{Sensitivity of 5-fold cross-validation performance to the feature-selection budget
  ($K = 10$--$72$) for *N$_2$, *N$_2$H, and *NH$_3$; selection performed via SelectKBest
  (F-regression). (a) CV $R^2$ and (b) CV MAE (eV) for the outlier-removed dataset
  ($|z| \leq 3$). (c) CV $R^2$ and (d) CV MAE (eV) for the unfiltered dataset.}
  \label{fig:fig7}
\end{figure}

The SelectKBest feature-selection budget was evaluated at $K = 10, 20, 30, 40, 50, 60$, and
$72$, for both the outlier-removed ($|z| \leq 3$) and unfiltered datasets
(Fig.~\ref{fig:fig7}). Within the outlier-removed dataset used throughout the remainder of
this section, cross-validation $R^2$ for *N$_2$ improved steadily with additional features,
from 0.618 at $K = 20$ to 0.657 at $K = 40$, before plateauing, and $K = 40$ was therefore
adopted for this adsorbate. *N$_2$H and *NH$_3$ showed no corresponding benefit (N$_2$H:
0.796 at both $K = 20$ and $K = 40$; NH$_3$: 0.697 at $K = 20$ declining to 0.672 at
$K = 40$), so the more compact $K = 20$ feature set was retained for both. This is in line
with the earlier result in this study that showed *N$_2$ adsorption is highly sensitive to
local adsorption geometry and benefits from a larger, more descriptive feature set, whereas
the adsorption energies of *N$_2$H and *NH$_3$ can be well captured by the compact
20-feature representation.

Comparing the outlier-removed and unfiltered datasets across this same sweep
(Fig.~\ref{fig:fig7}) shows that cross-validation $R^2$ is higher with outlier removal for
all three adsorbates at every value of $K$, with the improvement most significant for
*N$_2$ and smallest for *NH$_3$. For *N$_2$, the gap is the largest of the three and remains
essentially constant across the sweep (a gap of 0.169 at $K = 10$ to 0.136 at $K = 40$
between filtered and unfiltered datasets), indicating a persistent improvement from outlier
removal regardless of feature-set size. *N$_2$H shows a smaller, modestly narrowing gap
(0.118 at $K = 20$ to 0.087 at $K = 40$). *NH$_3$ shows a gap that narrows from 0.075 at
$K = 10$ to just 0.003 by $K = 40$, indicating that the cross-validation benefit of outlier
removal becomes minimal for this adsorbate once a sufficiently rich feature set is used.
Because outlier removal is never worse under cross-validation and is comparable-to-better on
the held-out test set, and because a small number of statistically extreme adsorption
energies can otherwise dominate feature selection (SelectKBest) and model fitting on a
dataset of this size, the $|z| \leq 3$ condition was retained as the primary
outlier-handling strategy for all results discussed in the main text. The corresponding
sweep on the unfiltered dataset, where $K = 40$ provides consistently strong performance
across all three adsorbates, is reported in full in the SI (Fig.~S2).

\FloatBarrier

\begin{figure}[htbp]
  \centering
  \includegraphics[width=\linewidth]{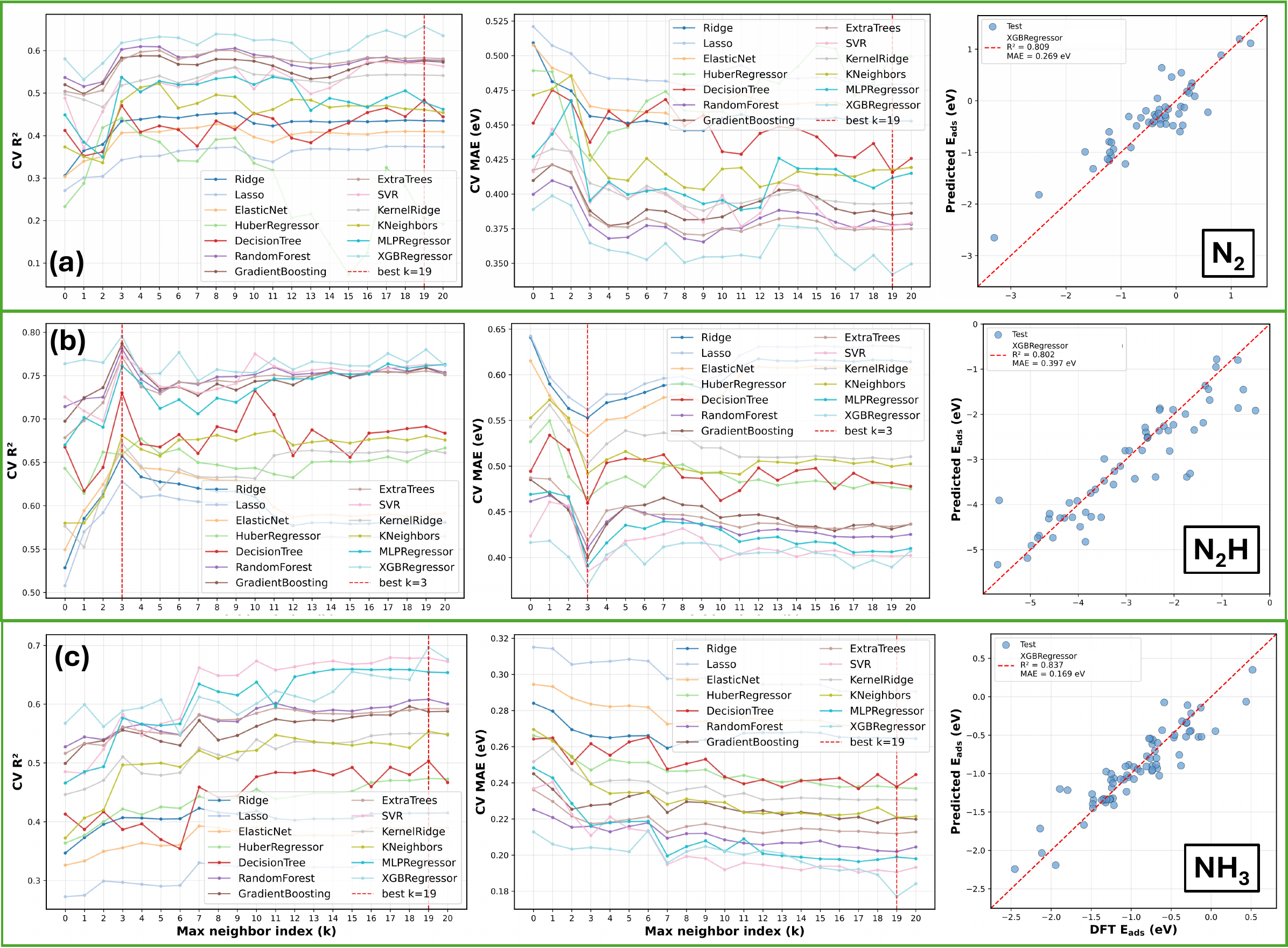}
  \caption{Cross-validation performance of all machine learning models as a function of the
  number of neighbor shells ($k$) considered in feature construction on the filtered dataset,
  with feature selection performed via SelectKBest (F-regression), and parity plots of the
  best-performing model evaluated on the sealed test set, for each adsorbate system:
  (a) N$_2$, (b) N$_2$H, and (c) NH$_3$. Left panels show the CV $R^2$ and middle panels
  show the CV MAE (eV) for all 13 models across $k = 0$ to $k = 20$, where $k = 0$
  corresponds to the binding site atom alone and $k = 20$ corresponds to the 21st nearest
  neighbor shell. The red dashed vertical line marks the optimal neighbor level selected by
  the highest CV $R^2$. Right panels show the parity plot of DFT-calculated versus
  ML-predicted adsorption energies on the sealed test set (20\% holdout) using the
  best-performing model at the optimal $k$.}
  \label{fig:fig8}
\end{figure}

Fig.~\ref{fig:fig8} summarizes the cross-validation $R^2$ and MAE of all models as a function
of neighbor shell size $k$ with the set feature-selection budget of $K = 40, 20, 20$ for
*N$_2$, *N$_2$H, and *NH$_3$, highlighting the best-performing model and optimal neighbor
level for each adsorbate. For N$_2$, XGB consistently outperformed all other algorithms
across the neighbor sweep, achieving its best cross-validation performance at $k = 19$
($n_0$ to $n_{19}$) with a CV $R^2$ of 0.657 and a CV MAE of 0.342~eV. Ensemble tree-based
methods such as RF (CV $R^2 = 0.610$ at $k = 4$) and GB (CV $R^2 = 0.588$ at $k = 4$) also
performed competitively, while linear models such as Ridge and Lasso remained substantially
weaker throughout the sweep (CV $R^2 < 0.46$), indicating that the relationship between the
feature set and N$_2$ adsorption energy is inherently nonlinear. On the held-out test set,
the best model achieved an $R^2$ of 0.809 and a MAE of 0.269~eV, demonstrating accurate
prediction of adsorption energies for held-out adsorption sites within the investigated
material space.

For N$_2$H, the CV performance landscape was markedly different. The best cross-validation
result was achieved by XGB at $k = 3$ ($n_0$ to $n_3$), with a CV $R^2$ of 0.796 and a CV
MAE of 0.369~eV. Notably, the performance of nearly all models peaks sharply at $k = 3$ and
then degrades or plateaus as more neighbor shells are added, with GB (CV $R^2 = 0.787$) and
RF (CV $R^2 = 0.781$) also achieving their optima at $k = 3$. This behavior suggests that
for N$_2$H, the local chemical environment within four neighbor shells already captures the
dominant structural and electronic factors governing adsorption, and that including more
distant atoms introduces noise rather than signal. Even linear models performed substantially
better for N$_2$H than for N$_2$, with Ridge and ENet reaching CV $R^2$ values above 0.65
at $k = 3$, suggesting a stronger linear component in the structure-property relationship for
this intermediate. On the test set, XGB yielded an $R^2$ of 0.802 and a MAE of 0.397~eV.

For NH$_3$, the CV performance trends were notably different from both N$_2$ and N$_2$H.
Rather than peaking early and stabilizing, the CV $R^2$ of the best-performing models
increased gradually and continuously with $k$, with XGB reaching its optimum at $k = 19$
($n_0$ to $n_{19}$) with a CV $R^2$ of 0.697 and a CV MAE of 0.177~eV. SVR emerged as a
strong competitor throughout the sweep, achieving a CV $R^2$ of 0.679 at $k = 17$ and
remaining among the top two models for almost all neighbor levels beyond $k = 5$. Despite the
lower CV $R^2$ compared to N$_2$H, the test set performance was the strongest of the three
molecules, with XGB achieving an $R^2$ of 0.837 and a MAE of 0.169~eV, reflecting the
narrower energy range and lower variance of NH$_3$ adsorption energies across the
intermetallic compound dataset.

Across all three adsorbates, XGB was the consistently best-performing algorithm, which can be
attributed to its ability to capture nonlinear feature interactions through gradient-boosted
decision trees while its internal regularization reduces overfitting on the relatively small
training sets.

\FloatBarrier

\begin{figure}[htbp]
  \centering
  \includegraphics[width=0.95\textwidth, height=0.65\textheight, keepaspectratio]{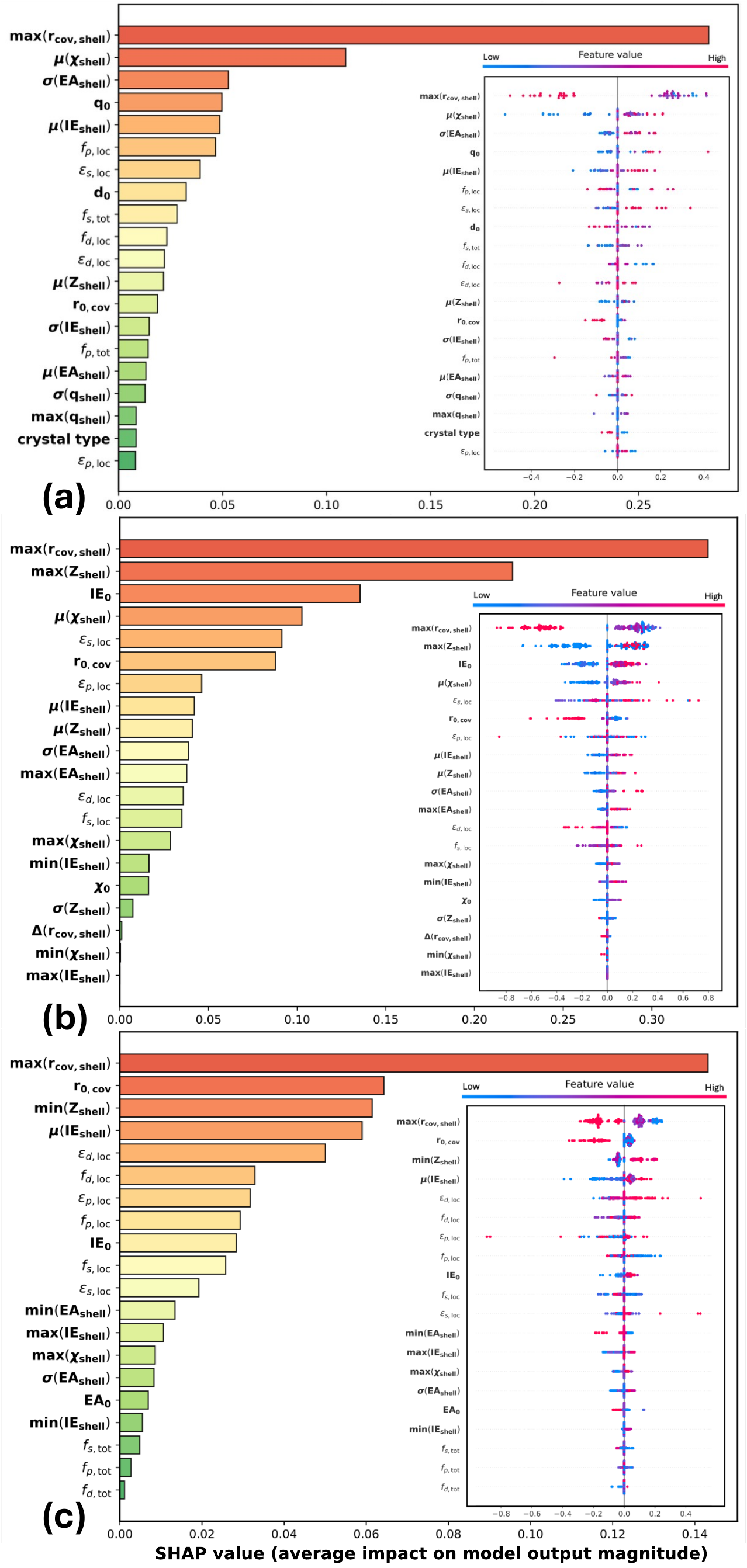}
  \begin{singlespace}
  \caption{SHAP feature importance analysis for the tuned best-performing models of each
  adsorbate system, using the $|z| \leq 3$ outlier-removed dataset with features selected via
  SelectKBest (F-regression) using $K=40$ for (a) N$_2$ and $K=20$ for (b) N$_2$H and
  (c) NH$_3$. Bars represent the mean absolute SHAP values of each descriptor,
  indicating its contribution to the model's prediction of adsorption energy. The descriptor
  set includes atomic properties of neighboring atoms (local), PDOS descriptors calculated for
  the full surface and neighboring shell, and shell statistics of intrinsic atomic properties
  such as mean, maximum, and minimum values computed across the neighbor shell (see Table~S3
  for full name, symbol, and definition of each descriptor). In the bottom-right corner of
  each panel, a beeswarm plot shows the directional impact of feature values on model
  predictions, with red indicating high feature values and blue indicating low values.}
  \end{singlespace}
  \label{fig:fig9}
\end{figure}

We assessed the importance of the selected features using SHAP bar plots for the
best-performing ML model corresponding to each adsorbate under the outlier-removed
($|z| \leq 3$) condition selected via SelectKBest (F-regression) (Fig.~\ref{fig:fig9}) and
RFE (Fig.~S3). In the bar plots, features are ranked vertically according to their mean
absolute SHAP contribution to the predicted adsorption energy, with the $x$-axis indicating
the magnitude of each feature's average impact. The accompanying beeswarm plots show the
distribution and directional effect of each feature across all training samples, where each
dot represents a single training sample colored by its feature value (red indicating high
values, blue indicating low values). The horizontal position of each dot corresponds to its
SHAP value showing how much that feature pushed the prediction toward more positive (weaker)
or more negative (stronger) adsorption energies.

A striking observation shared across all three adsorbates is the dominant role of
$\max(r_{\mathrm{cov,shell}})$, which is the maximum covalent radius across the neighbor
shell. This feature is ranked as the most important feature for N$_2$, N$_2$H, and NH$_3$.
This feature suggests that the presence of large-radius atoms in the neighbor shell,
regardless of which specific neighbor they are, is a fundamental structural factor controlling
adsorption strength across all three nitrogen-containing intermediates on intermetallic
surfaces. In the beeswarm plots, high values of $\max(r_{\mathrm{cov,shell}})$ are associated
with strongly negative SHAP values (red dots on the left), indicating that surfaces where the
neighbor shell contains at least one large-radius atom tend to bind all three intermediates
more strongly.

For N$_2$ adsorption, SHAP analysis at the optimal neighbor level of $k = 19$ (corresponding
to a neighbor shell extending up to the 20th nearest atom around the adsorption site) reveals
that beyond $\max(r_{\mathrm{cov,shell}})$ with a mean $|\mathrm{SHAP}|$ of 0.284, the mean
electronegativity of the shell ($\mu(\chi_\mathrm{shell})$) and the standard deviation of the
neighbor-shell electron affinity ($\sigma(\mathrm{EA}_\mathrm{shell})$) are the second and
third most important features, with mean $|\mathrm{SHAP}|$ values of 0.109 and 0.053~eV
respectively. The Bader charge of the first nearest neighbor ($q_0$) and the mean ionization
energy of the shell ($\mu(\mathrm{IE}_\mathrm{shell})$) also contribute meaningfully, followed
by the local p-band filling ($f_{p,\mathrm{loc}}$) and the local s-band center
($\varepsilon_{s,\mathrm{loc}}$). Notably, the d-band center and filling---which play the
central role in the classical d-band model---do not appear among the top features selected
for N$_2$, suggesting that for this dataset of intermetallic compounds, broader electronic
and structural descriptors of the local environment carry more predictive power than the
d-band alone. The beeswarm plot shows that low values of $\mu(\chi_\mathrm{shell})$
consistently push predictions to negative value, linking less electronegative neighbor
environments to stronger N$_2$ binding, while a high $\varepsilon_{s,\mathrm{loc}}$ value
pushes the adsorption energy to higher (more positive) values. For clarity, only the top 20
features used in the *N$_2$ model are shown; the remaining features had negligible mean
$|\mathrm{SHAP}|$ values ($\leq 0.008$, compared to a maximum of 0.284 for the top-ranked
feature) and are reported in full in Table~S9 in the SI. To confirm this ranking is not an
artifact of the feature-selection method, we repeated the SHAP analysis using RFE ($K = 40$,
$|z| \leq 3$) for *N$_2$ and obtained a similar top-feature ranking and $|\mathrm{SHAP}|$
values, since RFE and SelectKBest converge on the same 38-feature survivor set at this $K$.
This confirms that the dominance of $\max(r_{\mathrm{cov,shell}})$ and
$\mu(\chi_\mathrm{shell})$ over the d-band center for *N$_2$ is robust to the choice of
feature-selection algorithm, not an artifact of SelectKBest's F-regression scoring.

For N$_2$H adsorption at $k = 3$ (corresponding to a neighbor shell extending up to the 4th
nearest atom around the adsorption site), the SHAP analysis reveals a broader distribution of
important features compared to N$_2$. After $\max(r_{\mathrm{cov,shell}})$, the maximum
atomic number in the shell ($\max(Z_\mathrm{shell})$) and the first ionization energy of the
adsorbate's first nearest neighbor ($\mathrm{IE}_0$) emerge as the second and third most
important features with mean absolute SHAP values of 0.222 and 0.136~eV respectively,
followed by $\mu(\chi_\mathrm{shell})$, $\varepsilon_{s,\mathrm{loc}}$, $r_{0,\mathrm{cov}}$,
and $\varepsilon_{p,\mathrm{loc}}$. The beeswarm shows that high $\max(Z_\mathrm{shell})$
values (red dots) predominantly fall on the negative SHAP side, meaning that neighbor shells
containing heavier elements tend to promote stronger N$_2$H adsorption. The
$\varepsilon_{d,\mathrm{loc}}$ and $f_{s,\mathrm{loc}}$ also contribute at lower levels,
indicating that both local electronic structure and intrinsic atomic character of the binding
site jointly govern N$_2$H binding. To test the robustness of this ranking, we repeated the
SHAP analysis using RFE ($K = 20$, $|z| \leq 3$) for *N$_2$H. Unlike *N$_2$, the ranking is
not preserved: the local d-band center ($\varepsilon_{d,\mathrm{loc}}$) rises from a
lower-level contributor under SelectKBest to the second most important feature under RFE
(mean $|\mathrm{SHAP}| = 0.167$, behind only $\max(r_{\mathrm{cov,shell}})$). This indicates
that, unlike *N$_2$, the relative contribution of the d-band center for *N$_2$H should be
interpreted with caution, as it is more sensitive to the choice of feature-selection method
than the broader structural and s/p-band descriptors identified above.

For NH$_3$ adsorption at the optimal $k = 19$ (corresponding to a neighbor shell extending up
to the 20th nearest atom around the adsorption site), the feature importance landscape shifts
considerably toward a more even distribution among the top descriptors, compared to N$_2$ or
N$_2$H adsorption. After $\max(r_{\mathrm{cov,shell}})$, $r_{0,\mathrm{cov}}$ and
$\min(Z_\mathrm{shell})$ emerge as prominent contributors, followed closely by
$\mu(\mathrm{IE}_\mathrm{shell})$ and $\varepsilon_{d,\mathrm{loc}}$, marking the first
appearance of the local d-band center as a top feature across the three adsorbates. The local
d-band filling ($f_{d,\mathrm{loc}}$), local p-band center ($\varepsilon_{p,\mathrm{loc}}$),
local p-band filling ($f_{p,\mathrm{loc}}$), and $\mathrm{IE}_0$ all contribute at comparable
levels. The beeswarm plot for NH$_3$ shows that high $\max(r_{\mathrm{cov,shell}})$ values
remain strongly associated with more negative SHAP values. The total band fillings
($f_{s,\mathrm{tot}}$, $f_{p,\mathrm{tot}}$, $f_{d,\mathrm{tot}}$) appear in the bottom of
the NH$_3$ ranking. To test the robustness of this ranking, we repeated the SHAP analysis
using RFE ($K = 20$, $|z| \leq 3$) for *NH$_3$. The local d-band center remains among the
top contributors under RFE, in fact rising from the fifth to the second most important
feature (mean $|\mathrm{SHAP}| = 0.070$, up from 0.047 under SelectKBest) at an essentially
similar optimal neighbor shell ($k = 20$ under RFE versus $k = 19$ under SelectKBest). Unlike
*N$_2$ and *N$_2$H, the local d-band center is a consistently important contributor to
*NH$_3$ binding regardless of the feature-selection method used.

Fig.~S4 shows the importance of the selected features using SHAP bar plots for the
best-performing ML model corresponding to each adsorbate under the unfiltered condition with
features selected via SelectKBest (F-regression). Comparing the SHAP rankings between the
outlier-removed and unfiltered conditions shows that the overall picture is not dramatically
different: s- and p-orbital-derived descriptors remain among the top-ranked features in both
cases for all three adsorbates, and the leading structural descriptor, the maximum covalent
radius in the neighbor shell, stays prominent regardless of outlier treatment. The clearer
advantage of outlier removal shows up in the beeswarm plots rather than in the ranked feature
importance itself: however few the extreme adsorption-energy outliers are, their
disproportionate influence compresses and dilutes the visible relationship between each
feature's value and its effect on the predicted adsorption energy (an effect that is more
pronounced for *NH$_3$ and *N$_2$H) whereas after outlier removal this relationship becomes
markedly more distinguishable across the remaining features. This gain in interpretability
is the main reason the outlier-removed condition is retained as the main basis for
SHAP-based interpretation in this work.

\FloatBarrier

The IMCs in this study can be broadly classified into three groups: those containing p-block
elements (e.g., Al), transition metals with filled or nearly filled d-orbitals (e.g., Pt, Cu,
Zn), and transition metals with partially filled d-orbitals (e.g., Co, Ni, V, and Fe).
Generally, in traditional transition metal catalysis, d-orbitals have been known as the
primary contributors to adsorbate interaction~\cite{Yang2022}. The role of s- and p-orbitals
becomes increasingly significant in alloys containing p-block elements or transition metals
with filled or nearly filled d-orbitals. This trend aligns with findings in the study done by
Wu et al.~\cite{Wu2022}, on main-group single-atom catalysts (SACs), which demonstrates that
adsorption energies of some intermediates (e.g., NO) strongly correlate with s-band and
p-band centers for systems involving Al, Ga, or Mg. Therefore, the importance of s- and
p-orbitals identified by the SHAP analysis is not unexpected for alloy configurations
containing elements such as Al, Zn, Cu, and Pt, which are present in around 800 out of the
1200 configurations. Furthermore, the relatively narrow distribution of d-band descriptors
(e.g., $f_d$ and $\varepsilon_d$) shown in Fig.~\ref{fig:fig6} suggests limited variability
in purely d-derived features across these systems. As a result, ML models relying only on
d-band descriptors may struggle to capture subtle differences in adsorption energies, whereas
s- and p-orbital features show a broader distribution and offer complementary information.

\begin{figure}[htbp]
  \centering
  \includegraphics[width=\linewidth]{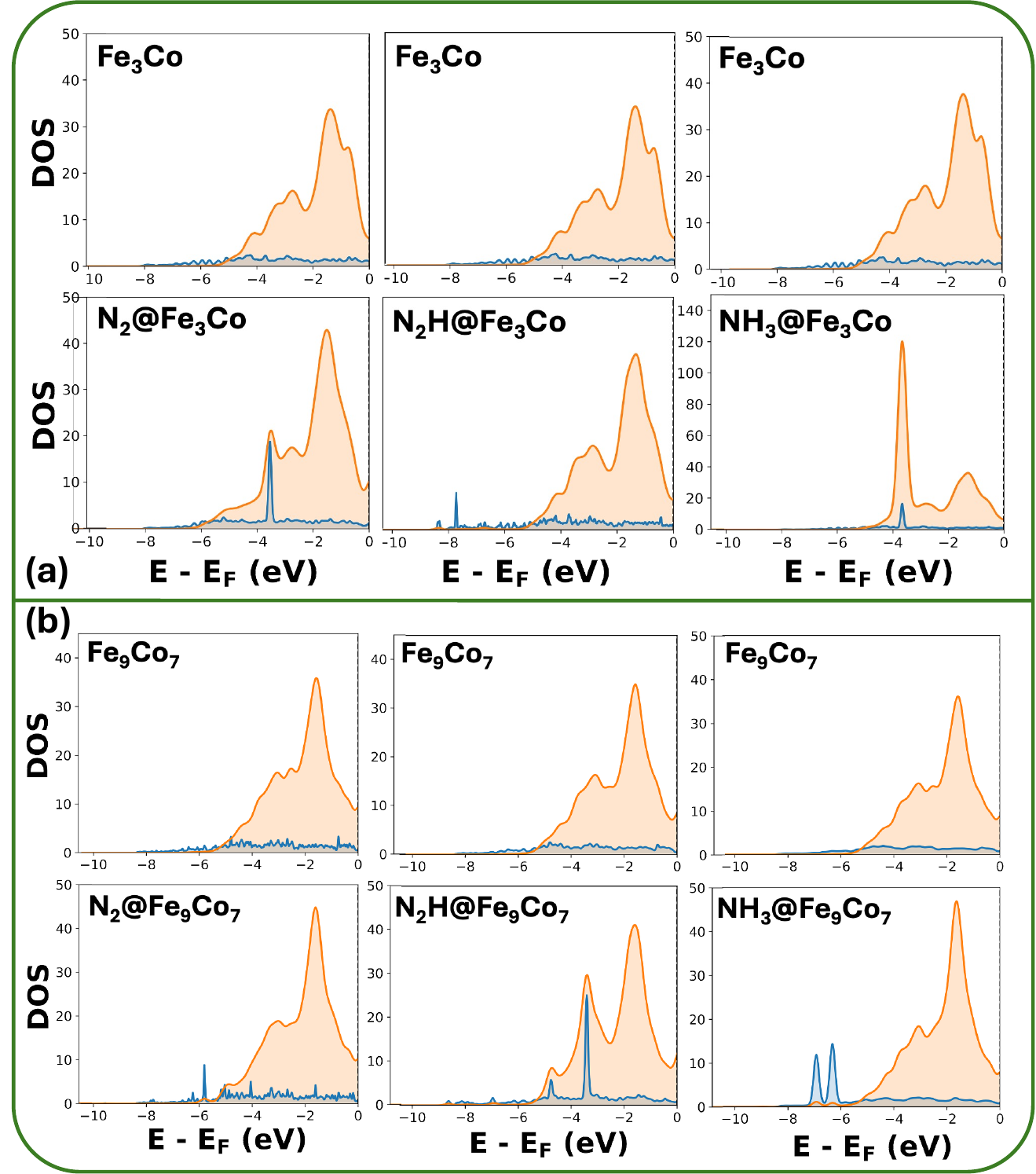}
  \caption{Atom-resolved projected density of states (PDOS) of metal atoms nearest to *N$_2$,
  *N$_2$H, and *NH$_3$ on (a) Fe$_3$Co and (b) Fe$_9$Co$_7$ surfaces. The top panels show the
  PDOS of surface metal atoms prior to adsorption, while the bottom panels correspond to the
  PDOS after adsorption of the respective intermediates. The blue and orange curves represent
  the sp- and d-orbital contributions, respectively.}
  \label{fig:fig10}
\end{figure}

To further investigate the role of s- and p-orbitals beyond systems containing p-block
elements or metals with filled or nearly filled d-band metals, we analyzed the projected
density of states (PDOS) of nearest-neighbor metal atoms to adsorbates in Fe$_3$Co and
Fe$_9$Co$_7$ (see Fig.~\ref{fig:fig10}). The top panels correspond to the pristine surfaces,
while the bottom panels represent the electronic structure after adsorption of N$_2$, N$_2$H,
and NH$_3$. Upon adsorption, noticeable changes occur in the sp states, including the
emergence of localized peaks and changes in DOS distribution. Importantly, in most cases, the
accentuated sp peaks appear in energy regions where d states are also present, indicating
that sp- and d-orbitals participate in the interaction. This co-localization and simultaneous
modification of sp and d states provide the evidence of hybridization between these orbitals
upon adsorption. Notably, this behavior observed in Fe$_3$Co and Fe$_9$Co$_7$, both of which
are composed of transition metals with partially filled d-orbitals, demonstrates that the
contribution of s- and p-orbitals is not limited to systems with filled or nearly filled
d-bands or to cases involving single-atom doping that alters the electronic
structure~\cite{Zhou2023ACS}. Instead, adsorption modifies the hybridized valence electronic
structure, where sp-orbitals actively participate in bonding alongside d states.

This pronounced sp-orbital contribution in d-block intermetallics can be explained by two
factors. First, unlike single-metal surfaces, where compact top-site binding
dominates~\cite{Xu2018,Hu2022}, IMCs favor side-on and tilted side-on N$_2$/N$_2$H
configurations spanning over multiple surface atoms (Fig.~\ref{fig:fig3}), and the more
spatially extended character of metal s- and p-states is generally understood to provide
better directional overlap with the adsorbate's $\pi^*$ and $\sigma$ orbitals across such
multi-atom sites than the more localized d-states. Second, in the classical two-step
(Newns--Anderson/Hammer--N{\o}rskov) picture of chemisorption, an adsorbate level hybridizes
with the broad continuum of metal sp-states, while the more localized d states provide
additional, strongly metal-dependent interactions. In the conventional d-band description,
the broad sp contribution is often treated as relatively similar across simple
transition-metal surfaces, with variations in adsorption primarily correlated with
characteristics of the d states~\cite{Hammer1995,Greeley2002}. Because the local coordination
environment in ordered intermetallics varies substantially from site to site, the local sp
electronic structure and its coupling to the adsorbate is no longer an invariant background
contribution but instead varies enough to carry independent predictive information. Together,
these factors indicate that adsorption in these transition-metal intermetallics is governed
by a coupled sp-d electronic structure whose local character varies significantly across
adsorption sites. The combination of extended sp states, multi-atom adsorption geometries,
and chemically distinct local environments provides a physical basis for why sp-band
features can contain independent information about adsorption energies and, consequently,
NRR activity. This provides a mechanistic interpretation of the statistical finding that
sp-band descriptors emerge as important predictive descriptors in these intermetallic
systems.

\FloatBarrier

\section{Conclusion}

In this study, we investigated NRR activity on IMCs by integrating high-throughput DFT
calculations with interpretable machine learning (ML) models. All adsorption sites across 47
bimetallic IMCs composed of Co, Ni, Al, Zn, V, Fe, Cu, and Pt were systematically explored,
generating a dataset of approximately 1,200 unique adsorption configurations. Among the
studied systems, Fe$_9$Co$_7$ and Fe$_3$Co exhibited the most balanced adsorption energies
across intermediates, indicating their potential as efficient NRR catalysts. Our analysis
shows that the traditional d-band center model alone cannot fully capture adsorption trends in
these multimetallic systems. Instead, both local and global s- and p-band characteristics
emerge as key descriptors governing adsorption strength, particularly for *N$_2$.
Furthermore, the local atomic environment plays a critical role, with *N$_2$ activation being
highly sensitive to neighboring atoms. Notably, accurate adsorption-energy predictions were
achieved using reduced sets of 38 physically meaningful features for *N$_2$ and 20 features
each for N$_2$H and NH$_3$, enabling the use of computationally efficient ML models. These
features include ten descriptors of the adsorbate's nearest-neighbor metal atom: its distance
from the adsorbate, atomic number, atomic mass, atomic and covalent radii, electronegativity,
electron affinity, first ionization energy, Bader charge, and atomic density. They also
include aggregated statistics of selected properties across the neighbor shell and local and
total PDOS descriptors, namely the s-, p-, and d-band centers and fillings. Together, these
features provide a compact, physically interpretable representation that captures the
electronic structure and local chemical environment of the adsorption site. Importantly, the
contribution of s- and p-orbitals is not limited to systems containing p-block elements or
nearly filled d-orbitals. Instead, analysis of Fe--Co systems reveals that in chemically
heterogeneous intermetallic surfaces, the local sp electronic structure becomes site-dependent
and participates in multi-atom adsorbate bonding alongside d states, making sp-band
characteristics informative descriptors of catalytic activity.

Collectively, this work challenges an exclusively d-band-centric view of adsorption in
intermetallic catalysts and establishes a compact, physically interpretable descriptor set
that governs NRR activity. Beyond identifying Fe$_9$Co$_7$ and Fe$_3$Co as promising,
non-noble catalysts for ammonia synthesis, this DFT--ML framework provides a transferable
methodology for decoding structure-activity relationships in compositionally complex
materials, opening a new path toward the rational design of next-generation catalysts for
sustainable nitrogen fixation.

\section*{Data and Code Availability}

The code used in this study, including all data processing, feature extraction, and machine
learning scripts, is publicly available on GitHub at \url{https://github.com/parezaei/NRR_Catalysis_ML}.

\section*{Acknowledgements}

K.K.G. acknowledges the support from Natural Sciences and Engineering Research Council of
Canada (NSERC) Discovery grant program [Funding Reference number RGPIN-2020-05924], Canada
Research Chair (CRC) program, the Canada Foundation for Innovation (CFI), Calcul Qu\'{e}bec
(\url{https://www.calculquebec.ca}), and the Digital Research Alliance of Canada
(\url{alliancecan.ca}). The authors thank Tanay Sahu, Satvik R Kashyap, and Subhomoy Gupta
for their contributions to data generation and Jonathan Shock for fruitful discussion.

\bibliographystyle{unsrtnat}
\bibliography{references}

\end{document}


\maketitle
\clearpage

\begin{figure}[htbp]
  \centering
  \includegraphics[width=\linewidth]{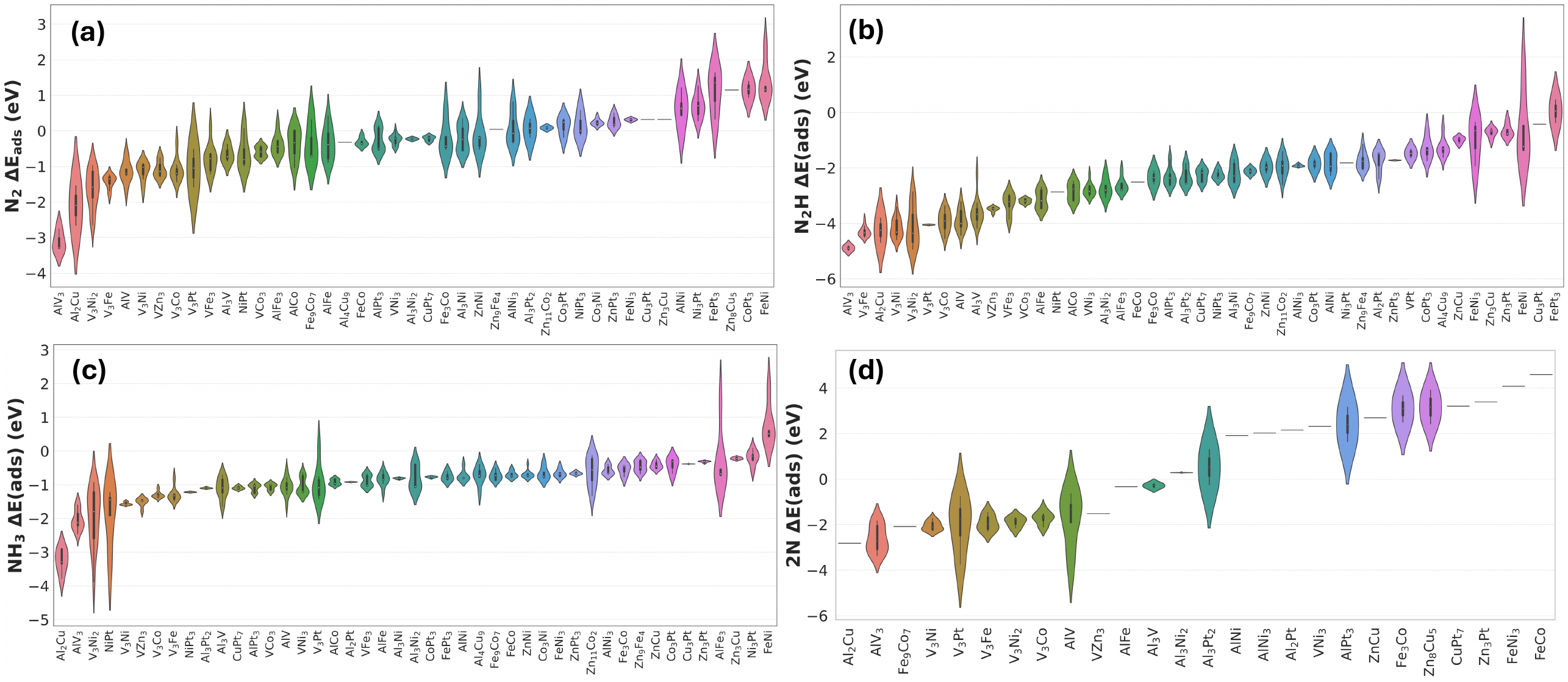}
  \begin{singlespace}
  \caption{Distributions (box plots) of adsorption energies across different adsorption sites
  in each IMC for molecular (a) N$_2$, (b) N$_2$H, (c) NH$_3$, and (d) dissociatively adsorbed
  N$_2$ (*2N).}
  \end{singlespace}
  \label{fig:figS1}
\end{figure}

\clearpage

\begin{figure}[htbp]
  \centering
  \includegraphics[width=\linewidth, height=0.7\textheight, keepaspectratio]{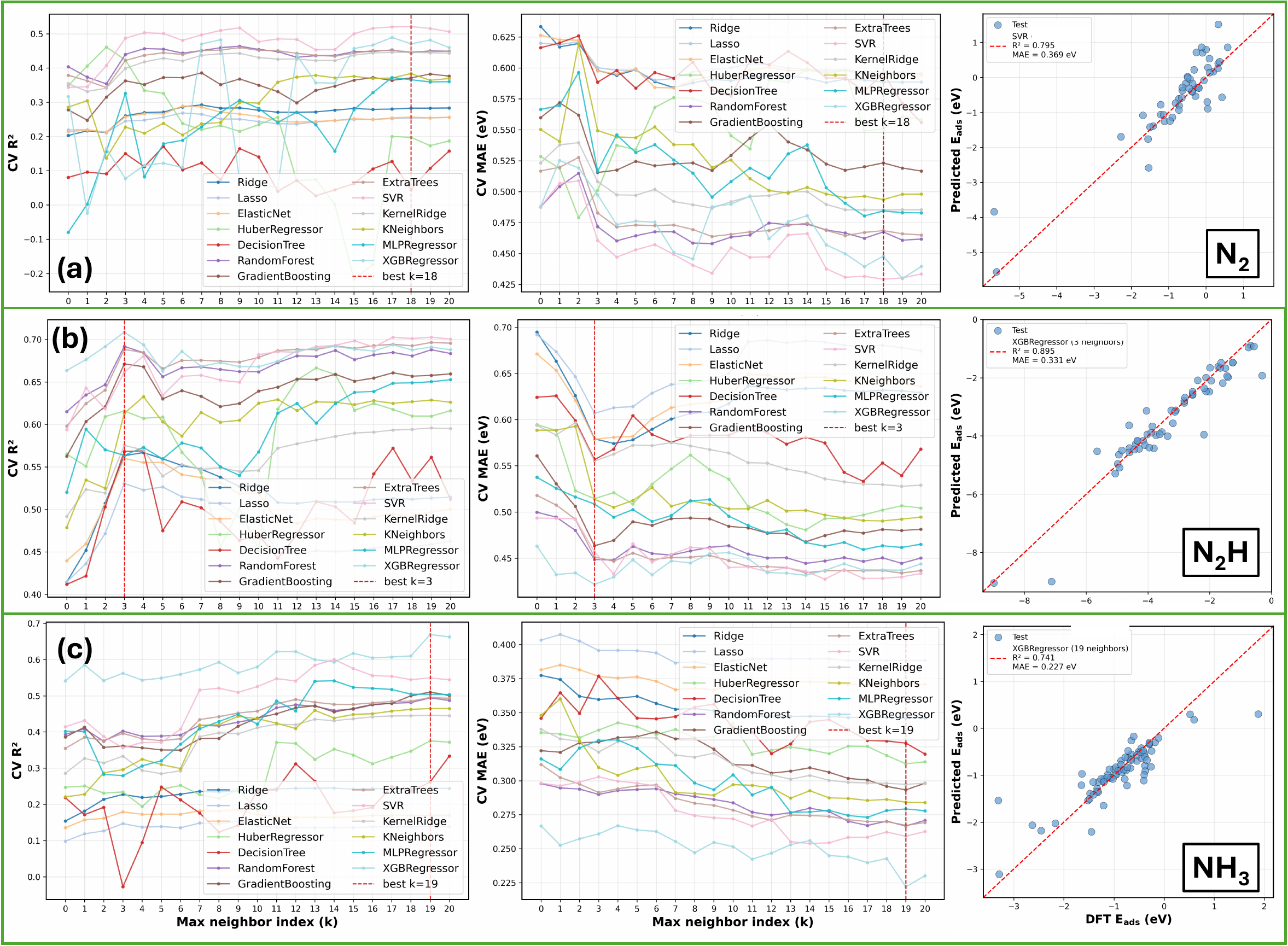}
  \begin{singlespace}
  \caption{Cross-validation performance of all machine learning models as a function of the
  number of neighbor shells ($k$) considered in feature construction on the unfiltered
  dataset, with feature selection performed via SelectKBest (F-regression) with a
  feature-selection budget of $K = 40$, and parity plots of the best-performing model
  evaluated on the sealed test set, for each adsorbate system: (a) N$_2$, (b) N$_2$H, and
  (c) NH$_3$. Left panels show the CV $R^2$ and middle panels show the CV MAE (eV) for all
  13 models across $k = 0$ to $k = 20$, where $k = 0$ corresponds to the binding site atom
  alone and $k = 20$ corresponds to the 21st nearest neighbor shell (see Table~S3 for full
  name, symbol, and definition of each descriptor). The red dashed vertical line marks the
  optimal neighbor level selected by the highest CV $R^2$. Right panels show the parity
  plot of DFT-calculated versus ML-predicted adsorption energies on the sealed test set
  (20\% holdout) using the best-performing model at the optimal $k$.}
  \end{singlespace}
  \label{fig:figS2}
\end{figure}

\clearpage

\begin{figure}[htbp]
  \centering
  \includegraphics[width=0.95\textwidth, height=0.65\textheight, keepaspectratio]{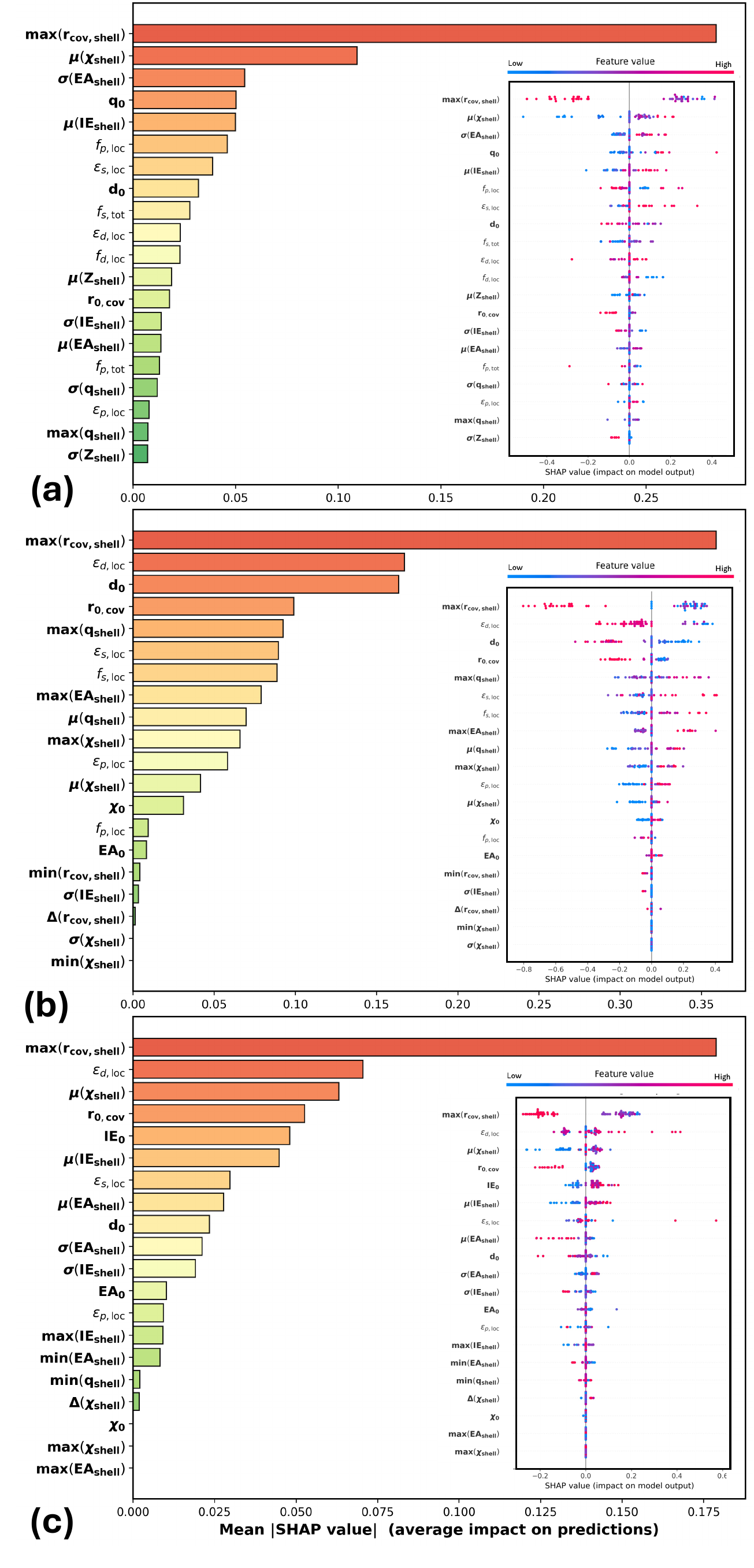}
  \begin{singlespace}
  \caption{SHAP feature importance analysis for the tuned best-performing models of each
  adsorbate system on the outlier-removed dataset ($|z| \leq 3$) with features selected via
  RFE, using a feature-selection budget of $K=40$ for N$_2$ and $K=20$ for N$_2$H and
  NH$_3$: (a) N$_2$, (b) N$_2$H, and (c) NH$_3$. Bars represent the mean absolute SHAP
  values of each descriptor, indicating its contribution to the model's prediction of
  adsorption energy. The descriptor set includes atomic properties of neighboring atoms
  (local), full surface and adsorbate's neighboring atoms PDOS features, and shell
  statistics of intrinsic atomic properties such as mean, maximum, and minimum values
  computed across the neighbor shell (see Table~S3 for full name, symbol, and definition of
  each descriptor). In the bottom-right corner of each panel, a beeswarm plot shows the
  directional impact of feature values on model predictions, with red indicating high
  feature values and blue indicating low values.}
  \end{singlespace}
  \label{fig:figS3}
\end{figure}

\clearpage

\begin{figure}[htbp]
  \centering
  \includegraphics[width=0.95\textwidth, height=0.65\textheight, keepaspectratio]{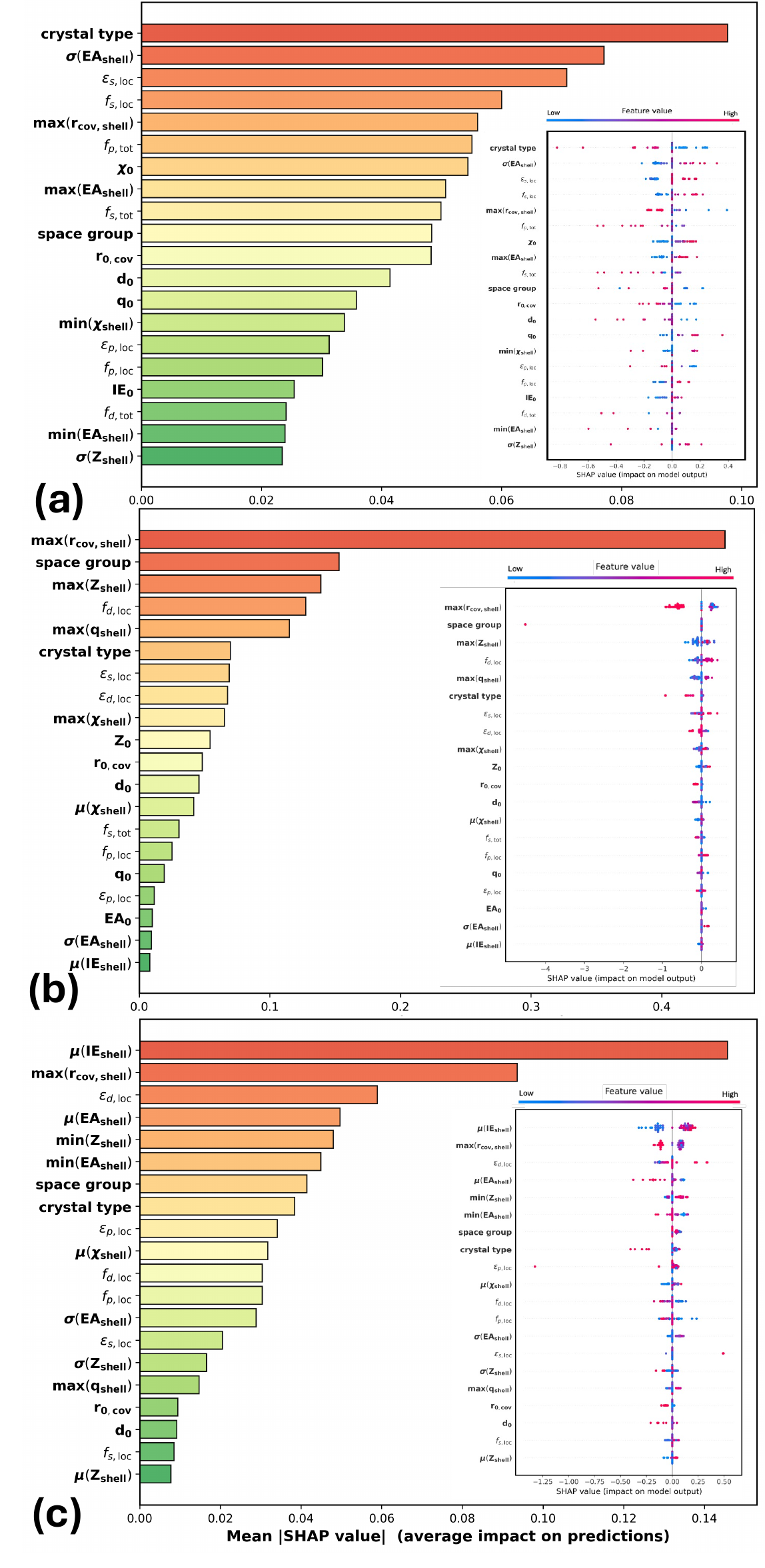}
  \begin{singlespace}
  \caption{SHAP feature importance analysis for the tuned best-performing models of each
  adsorbate system on the unfiltered dataset with features selected via SelectKBest
  (F-regression) and $K=40$ for all three adsorbates: (a) N$_2$, (b) N$_2$H, and (c) NH$_3$.
  Bars represent the mean absolute SHAP values of each descriptor, indicating its
  contribution to the model's prediction of adsorption energy. The descriptor set includes
  atomic properties of neighboring atoms (local), full surface and adsorbate's neighboring
  atoms PDOS features, and shell statistics of intrinsic atomic properties such as mean,
  maximum, and minimum values computed across the neighbor shell (see Table~S3 for full
  name, symbol, and definition of each descriptor). In the bottom-right corner of each
  panel, a beeswarm plot shows the directional impact of feature values on model
  predictions, with red indicating high feature values and blue indicating low values.}
  \end{singlespace}
  \label{fig:figS4}
\end{figure}

\clearpage

\begin{landscape}
\singlespacing
\footnotesize
\setlength{\LTcapwidth}{\linewidth}
\begin{longtable}{@{}llccrrrrrrccc@{}}
  \caption{Materials Project identifiers, crystal systems, bulk formation energies, and DFT
  model parameters for the 47 intermetallic compounds investigated. The model parameters
  include the selected surface plane, numbers of atoms in the unit cell, bulk supercell, and
  surface slab, lateral slab dimensions, and bulk magnetization. Crystal-system and
  bulk-formation-energy data were obtained from the Materials Project
  database~\cite{Horton2025}. The ``Bulk DFT magnetization'' values were obtained from our
  bulk DFT calculations and are included for comparison with values from the Materials
  Project~\cite{Horton2025} or available literature~\cite{Mondal2016,Paudyal2004,Parra1980};
  shaded rows correspond to IMCs containing at least one of the magnetic elements Fe, Co, or
  Ni.}
  \label{tab:tabS1}\\
  \toprule
  \textbf{Mat.} & \textbf{MP-ID} & \textbf{Cryst.\ Sys.} & \textbf{Bulk FE} &
  \textbf{Plane} & \textbf{$N_{\mathrm{uc}}$} & \textbf{Exp.} & \textbf{$N_{\mathrm{bulk}}$} &
  \textbf{$N_{\mathrm{surf}}$} & \textbf{Lat.\ $a$} & \textbf{Lat.\ $b$} &
  \textbf{Rep.\ Mag.} & \textbf{DFT Mag.} \\
   & & \textbf{(space grp.)} & \textbf{(eV/atom)} & & & \textbf{factor} & & &
  \textbf{(\AA)} & \textbf{(\AA)} & \textbf{($\mu_B$/f.u.)} & \textbf{($\mu_B$/f.u.)} \\
  \midrule
  \endfirsthead
  \toprule
  \textbf{Mat.} & \textbf{MP-ID} & \textbf{Cryst.\ Sys.} & \textbf{Bulk FE} &
  \textbf{Plane} & \textbf{$N_{\mathrm{uc}}$} & \textbf{Exp.} & \textbf{$N_{\mathrm{bulk}}$} &
  \textbf{$N_{\mathrm{surf}}$} & \textbf{Lat.\ $a$} & \textbf{Lat.\ $b$} &
  \textbf{Rep.\ Mag.} & \textbf{DFT Mag.} \\
   & & \textbf{(space grp.)} & \textbf{(eV/atom)} & & & \textbf{factor} & & &
  \textbf{(\AA)} & \textbf{(\AA)} & \textbf{($\mu_B$/f.u.)} & \textbf{($\mu_B$/f.u.)} \\
  \midrule
  \endhead
  \bottomrule
  \endfoot
  Al$_2$Cu   & mp-985806  & Cubic (Fm-3m)      & $-0.213$ & $(1,0,1)$     & 3  & 16 & 48  & 64  & 7.81  & 15.12 & --            & --    \\
  Al$_2$Pt   & mp-1502    & Cubic (Fm-3m)      & $-0.990$ & $(-1,1,-1)$   & 3  & 32 & 96  & 96  & 11.89 & 16.81 & --            & --    \\
  \rowcolor{magrow}
  Al$_3$Ni   & mp-622209  & Ortho.\ (Pnma)     & $-0.419$ & $(0,0,2)$     & 16 & 8  & 128 & 88  & 9.85  & 13.11 & 0.0 [1]       & 0.0   \\
  Al$_3$Ni$_2$ & mp-1057  & Trigonal (P-3m1)   & $-0.644$ & $(0,0,-1)$    & 5  & 8  & 40  & 160 & 15.85 & 15.85 & --            & --    \\
  Al$_3$Pt$_2$ & mp-10905 & Trigonal (P-3m1)   & $-1.103$ & $(0,0,-1)$    & 5  & 8  & 40  & 160 & 16.59 & 16.59 & --            & --    \\
  Al$_3$V    & mp-2554    & Tetrag.\ (I4/mmm)  & $-0.344$ & $(1,-1,2)$    & 4  & 16 & 64  & 128 & 10.68 & 19.79 & --            & --    \\
  Al$_4$Cu$_9$ & mp-593   & Cubic (P-43m)      & $-0.233$ & $(0,3,-3)$    & 52 & 1  & 52  & 72  & 8.73  & 12.34 & --            & --    \\
  \rowcolor{magrow}
  AlCo       & mp-284     & Cubic (Pm-3m)      & $-0.556$ & $(1,1,0)$     & 2  & 64 & 128 & 128 & 11.44 & 16.18 & 0.0 [1]       & 0.0   \\
  \rowcolor{magrow}
  AlFe       & mp-2658    & Cubic (Pm-3m)      & $-0.327$ & $(1,1,0)$     & 2  & 8  & 16  & 128 & 11.84 & 16.74 & 1.89 [2]      & 2.25  \\
  \rowcolor{magrow}
  AlFe$_3$   & mp-1183162 & Cubic (Fm-3m)      & $-0.199$ & $(1,1,-1)$    & 4  & 8  & 32  & 64  & 10.39 & 10.39 & 7.32 [1]      & 7.31  \\
  \rowcolor{magrow}
  AlNi       & mp-1487    & Cubic (Pm-3m)      & $-0.685$ & $(1,1,0)$     & 2  & 8  & 16  & 128 & 11.64 & 16.46 & 0.0 [1]       & 0.0   \\
  \rowcolor{magrow}
  AlNi$_3$   & mp-2593    & Cubic (Pm-3m)      & $-0.426$ & $(-1,1,1)$    & 4  & 8  & 32  & 64  & 10.09 & 10.09 & 0.92 [1]      & 1.23  \\
  AlPt$_3$   & mp-607111  & Tetrag.\ (P4/mbm)  & $-0.753$ & $(0,-2,2)$    & 16 & 8  & 128 & 120 & 10.83 & 19.64 & --            & --    \\
  AlV        & mp-1228001 & Ortho.\ (Cmmm)     & $-0.286$ & $(1,1,1)$     & 2  & 16 & 32  & 64  & 10.72 & 10.79 & --            & --    \\
  AlV$_3$    & mp-1387    & Cubic (Pm-3-n)     & $-0.169$ & $(1,0,2)$     & 8  & 8  & 64  & 164 & 9.58  & 21.43 & --            & --    \\
  \rowcolor{magrow}
  Co$_3$Ni   & mp-1183837 & Hexag.\ (P6$_3$/mmc) & $-0.023$ & $(0,0,2)$   & 8  & 8  & 64  & 64  & 9.87  & 9.87  & 6.08 [1]      & 5.76  \\
  \rowcolor{magrow}
  Co$_3$Pt   & mp-865193  & Hexag.\ (P6$_3$/mmc) & $-0.076$ & $(0,0,2)$   & 8  & 8  & 64  & 64  & 10.32 & 10.32 & 6.03 [1]      & 5.76  \\
  \rowcolor{magrow}
  CoPt$_3$   & mp-922     & Cubic (Pm-3m)      & $-0.048$ & $(1,1,-1)$    & 4  & 8  & 32  & 64  & 10.87 & 10.87 & 2.85 [1]      & 2.20  \\
  Cu$_3$Pt   & mp-12086   & Cubic (Pm-3m)      & $-0.160$ & $(1,1,-1)$    & 4  & 8  & 32  & 64  & 10.57 & 10.57 & --            & --    \\
  CuPt$_7$   & mp-12608   & Cubic (Fm-3m)      & $-0.106$ & $(2,2,-2)$    & 8  & 8  & 32  & 64  & 11.04 & 11.04 & --            & --    \\
  \rowcolor{magrow}
  Fe$_3$Co   & mp-601820  & Tetrag.\ (P4/mmm)  & $-0.042$ & $(-2,0,0)$    & 4  & 27 & 108 & 72  & 8.54  & 12.13 & 9.78 [1]      & 9.81  \\
  \rowcolor{magrow}
  Fe$_9$Co$_7$ & mp-601842 & Cubic (Fm-3m)     & $-0.057$ & $(0,2,2)$     & 16 & 8  & 128 & 128 & 11.35 & 16.05 & 38.45 [1]     & 34.84 \\
  \rowcolor{magrow}
  FeCo       & mp-2090    & Cubic (Fm-3m)      & $-0.055$ & $(1,-1,0)$    & 2  & 27 & 54  & 72  & 8.48  & 12.00 & 4.78 [1]      & 4.40  \\
  \rowcolor{magrow}
  FeNi       & mp-2213    & Tetrag.\ (P4/mmm)  & $-0.067$ & $(1,0,-1)$    & 4  & 12 & 48  & 72  & 7.67  & 12.91 & 3.54 [1]      & 3.58  \\
  \rowcolor{magrow}
  FeNi$_3$   & mp-1418    & Cubic (Pm-3m)      & $-0.090$ & $(-1,-1,1)$   & 4  & 8  & 32  & 64  & 10.02 & 10.02 & 5.28 [1]      & 4.78  \\
  \rowcolor{magrow}
  FePt$_3$   & mp-649     & Cubic (Pm-3m)      & $-0.174$ & $(1,1,-1)$    & 4  & 8  & 32  & 64  & 10.96 & 10.96 & 3.9--4.44 [3] & 3.20  \\
  \rowcolor{magrow}
  Ni$_3$Pt   & mp-12798   & Cubic (Pm-3m)      & $-0.069$ & $(-1,-1,-1)$  & 4  & 8  & 32  & 64  & 10.27 & 10.27 & 1.72--2.52 [3] & 1.87 \\
  \rowcolor{magrow}
  NiPt       & mp-945     & Tetrag.\ (P4/mmm)  & $-0.095$ & $(1,0,1)$     & 4  & 12 & 48  & 72  & 8.69  & 12.71 & 0.44 [3,4]    & 0.30  \\
  \rowcolor{magrow}
  NiPt$_3$   & mp-1186117 & Cubic (Pm-3m)      & $-0.063$ & $(1,1,-1)$    & 4  & 8  & 32  & 64  & 10.88 & 10.88 & 0.84--0.95 [4] & 0.94 \\
  \rowcolor{magrow}
  V$_3$Co    & mp-1585    & Cubic (Pm-3n)      & $-0.146$ & $(-2,0,0)$    & 8  & 8  & 64  & 60  & 9.27  & 9.27  & 0.0 [1]       & 0.0   \\
  \rowcolor{magrow}
  V$_3$Fe    & mp-1079399 & Cubic (Pm-3n)      & $-0.171$ & $(-2,0,0)$    & 8  & 8  & 64  & 60  & 9.27  & 9.27  & 0.0 [1]       & 0.0   \\
  \rowcolor{magrow}
  V$_3$Ni    & mp-7226    & Cubic (Pm-3n)      & $-0.154$ & $(-2,0,0)$    & 8  & 8  & 64  & 60  & 9.30  & 9.30  & 0.0 [1]       & 0.0   \\
  \rowcolor{magrow}
  V$_3$Ni$_2$ & mp-1216708 & Monoclinic (P2/m) & $-0.178$ & $(0,-2,0)$    & 30 & 1  & 30  & 66  & 11.44 & 11.57 & 0.0 [1]       & 0.21  \\
  V$_3$Pt    & mp-2211    & Cubic (Pm-3n)      & $-0.427$ & $(0,0,2)$     & 8  & 32 & 216 & 144 & 14.36 & 14.36 & --            & --    \\
  \rowcolor{magrow}
  VCo$_3$    & mp-542614  & Hexag.\ (P6$_3$/mmc) & $+0.011$ & $(0,0,-6)$  & 24 & 1  & 24  & 64  & 9.73  & 9.73  & 0.0 [1]       & 0.0   \\
  \rowcolor{magrow}
  VFe$_3$    & mp-866134  & Cubic (Fm-3m)      & $-0.128$ & $(2,-2,0)$    & 4  & 32 & 128 & 128 & 11.40 & 16.12 & 5.49 [1]      & 5.01  \\
  \rowcolor{magrow}
  VNi$_3$    & mp-171     & Tetrag.\ (I4/mmm)  & $-0.216$ & $(-1,1,-2)$   & 4  & 16 & 64  & 128 & 9.89  & 17.79 & 0.0 [1]       & 0.0   \\
  VPt        & mp-2678    & Tetrag.\ (P4/mmm)  & $-0.550$ & $(0,0,-1)$    & 4  & 12 & 48  & 45  & 8.20  & 8.20  & --            & --    \\
  VZn$_3$    & mp-11578   & Cubic (Pm-3m)      & $-0.074$ & $(-1,1,1)$    & 4  & 8  & 32  & 64  & 10.81 & 10.81 & --            & --    \\
  \rowcolor{magrow}
  Zn$_{11}$Co$_2$ & mp-1192361 & Cubic (I-43m) & $-0.056$ & $(-3,0,3)$    & 26 & 2  & 52  & 74  & 8.96  & 12.66 & 0.0 [1]       & 0.0   \\
  Zn$_3$Cu   & mp-972042  & Hexag.\ (P6$_3$/mmc) & $-0.081$ & $(0,0,-2)$  & 8  & 8  & 64  & 80  & 10.87 & 10.87 & --            & --    \\
  Zn$_3$Pt   & mp-865355  & Tetrag.\ (I4/mmm)  & $-0.400$ & $(-1,1,2)$    & 4  & 16 & 64  & 128 & 10.62 & 19.45 & --            & --    \\
  Zn$_8$Cu$_5$ & mp-1368  & Cubic (I-43m)      & $-0.108$ & $(0,3,3)$     & 26 & 2  & 52  & 70  & 8.85  & 12.52 & --            & --    \\
  \rowcolor{magrow}
  Zn$_9$Fe$_4$ & mp-1207450 & Cubic (I-43m)    & $-0.043$ & $(3,3,0)$     & 26 & 2  & 52  & 76  & 8.90  & 12.59 & 10.20 [1]     & 4.91  \\
  ZnCu       & mp-987     & Cubic (Pm-3m)      & $-0.091$ & $(1,-1,0)$    & 2  & 27 & 54  & 72  & 8.90  & 12.58 & --            & --    \\
  \rowcolor{magrow}
  ZnNi       & mp-429     & Tetrag.\ (P4/mmm)  & $-0.255$ & $(0,1,1)$     & 4  & 12 & 48  & 72  & 8.19  & 12.72 & 0.0 [1]       & 0.0   \\
  ZnPt$_3$   & mp-30856   & Cubic (Pm-3m)      & $-0.313$ & $(1,1,-1)$    & 4  & 8  & 32  & 64  & 11.04 & 11.04 & --            & --    \\
\end{longtable}
\footnotesize{References for reported bulk magnetization: [1] Materials Project
database~\cite{Horton2025}; [2] Mondal et al.~\cite{Mondal2016}; [3] Paudyal et
al.~\cite{Paudyal2004}; [4] Parra and Cable~\cite{Parra1980}.}
\normalsize
\end{landscape}

\clearpage

\singlespacing
\begin{table}[htbp]
  \centering
  \caption{Properties of selected elements considered in this study. The number of
  d-electrons, Pauling electronegativity, and valence electron configuration are listed.\\}
  \label{tab:tabS2}
  \begin{tabular}{lccc}
    \toprule
    \textbf{Element} & \textbf{Number of d-electrons} & \textbf{Electronegativity} &
    \textbf{Valence Electron Config.} \\
    \midrule
    Co & 7 & 1.88 & $3d^7\,4s^2$ \\
    Ni & 8 & 1.91 & $3d^8\,4s^2$ \\
    Al & 0 & 1.61 & $3s^2\,3p^1$ \\
    Zn & 10 & 1.65 & $3d^{10}\,4s^2$ \\
    V  & 3 & 1.63 & $3d^3\,4s^2$ \\
    Fe & 6 & 1.83 & $3d^6\,4s^2$ \\
    Cu & 10 & 1.90 & $3d^{10}\,4s^1$ \\
    Pt & 9 & 2.28 & $5d^9\,6s^1$ \\
    \bottomrule
  \end{tabular}
\end{table}
\doublespacing

\clearpage

\singlespacing
\setlength{\LTcapwidth}{\linewidth}
\begin{longtable}{lp{11cm}}
  \caption{Definitions and symbols of the machine learning descriptors used to predict
  adsorption energies, comprising global structural, first-nearest-neighbor ($n_0$),
  shell-aggregated, and projected density-of-states (PDOS) descriptors.}
  \label{tab:tabS3}\\
  \toprule
  \textbf{Symbol} & \textbf{Description} \\
  \midrule
  \endfirsthead
  \toprule
  \textbf{Symbol} & \textbf{Description} \\
  \midrule
  \endhead
  \bottomrule
  \endfoot
  \multicolumn{2}{l}{\textit{Global structural descriptors}} \\
  Crystal type & Crystal structure type of the bulk intermetallic compound obtained from the
  Materials Project database (Table~S1). \\
  Space group & Crystallographic space group of the bulk structure, obtained from the
  Materials Project database (Table~S1). \\[4pt]
  \multicolumn{2}{l}{\textit{Adsorbate nearest-neighbor atom ($n_0$) properties}} \\
  $Z_0$ & Atomic number of the nearest-neighbor atom ($n_0$) to the adsorbate. \\
  $M_0$ & Atomic mass of the nearest-neighbor atom ($n_0$). \\
  $\mathrm{EA}_0$ & Electron affinity of the nearest-neighbor atom ($n_0$). \\
  $\mathrm{IE}_0$ & First ionization energy of the nearest-neighbor atom ($n_0$). \\
  $r_0$ & Atomic radius of the nearest-neighbor atom ($n_0$). \\
  $r_{0,\mathrm{cov}}$ & Covalent radius of the nearest-neighbor atom ($n_0$). \\
  $\rho_0$ & Atomic density of the nearest-neighbor atom ($n_0$). \\
  $d_0$ & Distance between the nearest-neighbor atom ($n_0$) and the adsorption site. \\
  $q_0$ & Bader charge of the nearest-neighbor atom ($n_0$), obtained from DFT
  calculations. \\
  $\chi_0$ & Pauling electronegativity of the nearest-neighbor atom ($n_0$). \\[4pt]
  \multicolumn{2}{l}{\textit{Shell-aggregated statistics (computed over the neighbor shell
  $n_0$ to $n_k$)}} \\
  \multicolumn{2}{l}{\textit{Pauling electronegativity ($\chi$)}} \\
  $\mu(\chi_{\mathrm{shell}})$ & Arithmetic mean of the electronegativity across the
  neighbor shell ($n_0$ to $n_k$). \\
  $\sigma(\chi_{\mathrm{shell}})$ & Standard deviation of the electronegativity across the
  neighbor shell. \\
  $\min(\chi_{\mathrm{shell}})$ & Minimum electronegativity among the atoms in the neighbor
  shell. \\
  $\max(\chi_{\mathrm{shell}})$ & Maximum electronegativity among the atoms in the neighbor
  shell. \\
  $\Delta(\chi_{\mathrm{shell}})$ & Range (maximum $-$ minimum) of the electronegativity
  across the neighbor shell. \\
  $\mu_w(\chi_{\mathrm{shell}})$ & Distance-weighted mean of the electronegativity across
  the neighbor shell, with closer neighbors weighted more heavily (Eq.~S4). \\[4pt]
  \multicolumn{2}{l}{\textit{Atomic number ($Z$)}} \\
  $\mu(Z_{\mathrm{shell}})$ & Arithmetic mean of the atomic number across the neighbor
  shell ($n_0$ to $n_k$). \\
  $\sigma(Z_{\mathrm{shell}})$ & Standard deviation of the atomic number across the
  neighbor shell. \\
  $\min(Z_{\mathrm{shell}})$ & Minimum atomic number among the atoms in the neighbor
  shell. \\
  $\max(Z_{\mathrm{shell}})$ & Maximum atomic number among the atoms in the neighbor
  shell. \\
  $\Delta(Z_{\mathrm{shell}})$ & Range (maximum $-$ minimum) of the atomic number across
  the neighbor shell. \\
  $\mu_w(Z_{\mathrm{shell}})$ & Distance-weighted mean of the atomic number across the
  neighbor shell, with closer neighbors weighted more heavily (Eq.~S4). \\[4pt]
  \multicolumn{2}{l}{\textit{Covalent radius ($r_{\mathrm{cov}}$)}} \\
  $\mu(r_{\mathrm{cov,shell}})$ & Arithmetic mean of the covalent radius across the
  neighbor shell ($n_0$ to $n_k$). \\
  $\sigma(r_{\mathrm{cov,shell}})$ & Standard deviation of the covalent radius across the
  neighbor shell. \\
  $\min(r_{\mathrm{cov,shell}})$ & Minimum covalent radius among the atoms in the neighbor
  shell. \\
  $\max(r_{\mathrm{cov,shell}})$ & Maximum covalent radius among the atoms in the neighbor
  shell. \\
  $\Delta(r_{\mathrm{cov,shell}})$ & Range (maximum $-$ minimum) of the covalent radius
  across the neighbor shell. \\
  $\mu_w(r_{\mathrm{cov,shell}})$ & Distance-weighted mean of the covalent radius across
  the neighbor shell, with closer neighbors weighted more heavily (Eq.~S4). \\[4pt]
  \multicolumn{2}{l}{\textit{Electron affinity (EA)}} \\
  $\mu(\mathrm{EA}_{\mathrm{shell}})$ & Arithmetic mean of the electron affinity across the
  neighbor shell ($n_0$ to $n_k$). \\
  $\sigma(\mathrm{EA}_{\mathrm{shell}})$ & Standard deviation of the electron affinity
  across the neighbor shell. \\
  $\min(\mathrm{EA}_{\mathrm{shell}})$ & Minimum electron affinity among the atoms in the
  neighbor shell. \\
  $\max(\mathrm{EA}_{\mathrm{shell}})$ & Maximum electron affinity among the atoms in the
  neighbor shell. \\
  $\Delta(\mathrm{EA}_{\mathrm{shell}})$ & Range (maximum $-$ minimum) of the electron
  affinity across the neighbor shell. \\
  $\mu_w(\mathrm{EA}_{\mathrm{shell}})$ & Distance-weighted mean of the electron affinity
  across the neighbor shell, with closer neighbors weighted more heavily (Eq.~S4). \\[4pt]
  \multicolumn{2}{l}{\textit{Bader charge ($q$)}} \\
  $\mu(q_{\mathrm{shell}})$ & Arithmetic mean of the Bader charge across the neighbor
  shell ($n_0$ to $n_k$). \\
  $\sigma(q_{\mathrm{shell}})$ & Standard deviation of the Bader charge across the neighbor
  shell. \\
  $\min(q_{\mathrm{shell}})$ & Minimum Bader charge among the atoms in the neighbor
  shell. \\
  $\max(q_{\mathrm{shell}})$ & Maximum Bader charge among the atoms in the neighbor
  shell. \\
  $\Delta(q_{\mathrm{shell}})$ & Range (maximum $-$ minimum) of the Bader charge across the
  neighbor shell. \\
  $\mu_w(q_{\mathrm{shell}})$ & Distance-weighted mean of the Bader charge across the
  neighbor shell, with closer neighbors weighted more heavily (Eq.~S4). \\[4pt]
  \multicolumn{2}{l}{\textit{Atomic density ($\rho$)}} \\
  $\mu(\rho_{\mathrm{shell}})$ & Arithmetic mean of the atomic density across the neighbor
  shell ($n_0$ to $n_k$). \\
  $\sigma(\rho_{\mathrm{shell}})$ & Standard deviation of the atomic density across the
  neighbor shell. \\
  $\min(\rho_{\mathrm{shell}})$ & Minimum atomic density among the atoms in the neighbor
  shell. \\
  $\max(\rho_{\mathrm{shell}})$ & Maximum atomic density among the atoms in the neighbor
  shell. \\
  $\Delta(\rho_{\mathrm{shell}})$ & Range (maximum $-$ minimum) of the atomic density
  across the neighbor shell. \\
  $\mu_w(\rho_{\mathrm{shell}})$ & Distance-weighted mean of the atomic density across the
  neighbor shell, with closer neighbors weighted more heavily (Eq.~S4). \\[4pt]
  \multicolumn{2}{l}{\textit{Atomic mass ($M$)}} \\
  $\mu(M_{\mathrm{shell}})$ & Arithmetic mean of the atomic mass across the neighbor shell
  ($n_0$ to $n_k$). \\
  $\sigma(M_{\mathrm{shell}})$ & Standard deviation of the atomic mass across the neighbor
  shell. \\
  $\min(M_{\mathrm{shell}})$ & Minimum atomic mass among the atoms in the neighbor
  shell. \\
  $\max(M_{\mathrm{shell}})$ & Maximum atomic mass among the atoms in the neighbor
  shell. \\
  $\Delta(M_{\mathrm{shell}})$ & Range (maximum $-$ minimum) of the atomic mass across the
  neighbor shell. \\
  $\mu_w(M_{\mathrm{shell}})$ & Distance-weighted mean of the atomic mass across the
  neighbor shell, with closer neighbors weighted more heavily (Eq.~S4). \\[4pt]
  \multicolumn{2}{l}{\textit{First ionization energy (IE)}} \\
  $\mu(\mathrm{IE}_{\mathrm{shell}})$ & Arithmetic mean of the first ionization energy
  across the neighbor shell ($n_0$ to $n_k$). \\
  $\sigma(\mathrm{IE}_{\mathrm{shell}})$ & Standard deviation of the first ionization
  energy across the neighbor shell. \\
  $\min(\mathrm{IE}_{\mathrm{shell}})$ & Minimum first ionization energy among the atoms in
  the neighbor shell. \\
  $\max(\mathrm{IE}_{\mathrm{shell}})$ & Maximum first ionization energy among the atoms in
  the neighbor shell. \\
  $\Delta(\mathrm{IE}_{\mathrm{shell}})$ & Range (maximum $-$ minimum) of the first
  ionization energy across the neighbor shell. \\
  $\mu_w(\mathrm{IE}_{\mathrm{shell}})$ & Distance-weighted mean of the first ionization
  energy across the neighbor shell, with closer neighbors weighted more heavily
  (Eq.~S4). \\[4pt]
  \multicolumn{2}{l}{\textit{Projected density-of-states (PDOS) descriptors (computed over
  the neighbor shell $n_0$ to $n_k$)}} \\
  \multicolumn{2}{l}{\textit{s-band}} \\
  $\varepsilon_{s,\mathrm{local}}$ & s-band center of the neighboring atoms around the
  adsorption site (Eq.~S2). \\
  $\varepsilon_{s,\mathrm{total}}$ & s-band center of the whole structure (Eq.~S2). \\
  $f_{s,\mathrm{local}}$ & s-band filling of the neighboring atoms around the adsorption
  site (Eq.~S3). \\
  $f_{s,\mathrm{total}}$ & s-band filling of the whole structure (Eq.~S3). \\[2pt]
  \multicolumn{2}{l}{\textit{p-band}} \\
  $\varepsilon_{p,\mathrm{local}}$ & p-band center of the neighboring atoms around the
  adsorption site (Eq.~S2). \\
  $\varepsilon_{p,\mathrm{total}}$ & p-band center of the whole structure (Eq.~S2). \\
  $f_{p,\mathrm{local}}$ & p-band filling of the neighboring atoms around the adsorption
  site (Eq.~S3). \\
  $f_{p,\mathrm{total}}$ & p-band filling of the whole structure (Eq.~S3). \\[2pt]
  \multicolumn{2}{l}{\textit{d-band}} \\
  $\varepsilon_{d,\mathrm{local}}$ & d-band center of the neighboring atoms around the
  adsorption site (Eq.~S2). \\
  $\varepsilon_{d,\mathrm{total}}$ & d-band center of the whole structure (Eq.~S2). \\
  $f_{d,\mathrm{local}}$ & d-band filling of the neighboring atoms around the adsorption
  site (Eq.~S3). \\
  $f_{d,\mathrm{total}}$ & d-band filling of the whole structure (Eq.~S3). \\
\end{longtable}
\doublespacing

\clearpage

\singlespacing
\begin{table}[htbp]
  \centering
  \caption{Hyperparameters of the machine learning algorithms (MLAs) considered in this study.
  The models include Ridge Regression (Ridge), Lasso Regression (Lasso), Elastic Net (ENet),
  Huber Regressor (Huber), Decision Tree Regressor (DT), Random Forest Regressor (RF),
  Gradient Boosting Regressor (GB), Extra Trees Regressor (ET), Support Vector Regressor (SVR),
  Kernel Ridge Regressor (KRR), k-Nearest Neighbors Regressor (KNN), Extreme Gradient
  Boosting Regressor (XGB), and Multi-Layer Perceptron (MLP). For tree-based models, max depth
  $=$ None indicates no restriction on tree growth.\\}
  \label{tab:tabS4}
  \begin{tabular}{lp{7cm}l}
    \toprule
    \textbf{ML model} & \textbf{Hyperparameters} & \textbf{Range} \\
    \midrule
    Ridge  & $\alpha$ (alpha) & 100 \\
    Lasso  & $\alpha$ (alpha), max iterations & 0.1, 5000 \\
    ENet   & $\alpha$ (alpha), l1 ratio, max iterations & 0.1, 0.5, 5000 \\
    Huber  & max iterations & 300 \\
    DT     & max depth, min samples leaf & 5, 4 \\
    RF     & n estimators, max depth, max features, min samples leaf & 100, 8, 0.3, 4 \\
    GB     & n estimators, max depth, learning rate, min samples leaf, subsample
           & 100, 2, 0.05, 5, 0.6 \\
    ET     & n estimators, max depth, max features, min samples leaf & 100, 8, 0.3, 4 \\
    SVR    & $C$, $\varepsilon$ (epsilon), kernel & 10, 0.1, RBF \\
    KRR    & $\alpha$ (alpha), kernel & 1.0, RBF \\
    KNN    & n neighbors & 5 \\
    XGB    & n estimators, max depth, learning rate, subsample, colsample bytree
           & 200, 4, 0.05, 0.8, 0.8 \\
    MLP    & hidden layer sizes, max iterations & (128, 64), 500 \\
    \bottomrule
  \end{tabular}
\end{table}
\doublespacing

\clearpage

\begin{landscape}
\singlespacing
\tiny
\setlength{\tabcolsep}{3.5pt}
\setlength{\LTcapwidth}{\linewidth}
\begin{longtable}{lcccccccccccc}
  \caption{\normalsize Comparison of recursive feature elimination (RFE) and SelectKBest with
  F-regression for *N$_2$, *N$_2$H, and *NH$_3$ adsorption-energy prediction under the
  outlier-removed ($|z| \leq 3$) and unfiltered conditions. Results are reported for
  feature-selection budgets of $K = 10, 20, 30, 40, 50, 60$, and $72$. ``Features used''
  denotes the actual number of features retained after variance and correlation filtering
  and is therefore sometimes smaller than the requested $K$. For each feature-selection
  method, the table reports the optimal neighbor level, best-performing model, repeated
  fivefold CV $R^2$, and CV MAE (eV). The final column gives the difference in CV
  performance.}
  \label{tab:tabS5}\\
  \toprule
  \textbf{Mol.} & \textbf{Outlier condition} & \textbf{$K$} & \textbf{Feat.\ used} &
  \textbf{RFE shell} & \textbf{RFE best model} & \textbf{RFE CV $R^2$} & \textbf{RFE CV MAE} &
  \textbf{SKB shell} & \textbf{SKB best model} & \textbf{SKB CV $R^2$} & \textbf{SKB CV MAE} &
  \textbf{$\Delta$ CV $R^2$ (RFE$-$SKB)} \\
  \midrule
  \endfirsthead
  \toprule
  \textbf{Mol.} & \textbf{Outlier condition} & \textbf{$K$} & \textbf{Feat.\ used} &
  \textbf{RFE shell} & \textbf{RFE best model} & \textbf{RFE CV $R^2$} & \textbf{RFE CV MAE} &
  \textbf{SKB shell} & \textbf{SKB best model} & \textbf{SKB CV $R^2$} & \textbf{SKB CV MAE} &
  \textbf{$\Delta$ CV $R^2$ (RFE$-$SKB)} \\
  \midrule
  \endhead
  \bottomrule
  \endfoot
  \multirow{7}{*}{*N$_2$}
  & \multirow{7}{*}{\shortstack{$|z|\leq3$ \\ (outlier-removed)}}
  & 10 & 10 & 8  & XGB & 0.6023 & 0.3756 & 7  & XGB & 0.6373 & 0.3653 & $-0.0350$ \\
  & & 20 & 20 & 5  & XGB & 0.6349 & 0.3589 & 10 & XGB & 0.6181 & 0.3578 & $+0.0169$ \\
  & & 30 & 30 & 17 & XGB & 0.6534 & 0.3428 & 17 & XGB & 0.6462 & 0.3482 & $+0.0073$ \\
  & & 40 & 38 & 19 & XGB & 0.6567 & 0.3419 & 19 & XGB & 0.6567 & 0.3419 & $+0.0000$ \\
  & & 50 & 38 & 19 & XGB & 0.6567 & 0.3419 & 19 & XGB & 0.6567 & 0.3419 & $+0.0000$ \\
  & & 60 & 38 & 19 & XGB & 0.6567 & 0.3419 & 19 & XGB & 0.6567 & 0.3419 & $+0.0000$ \\
  & & 72 & 38 & 19 & XGB & 0.6567 & 0.3419 & 19 & XGB & 0.6567 & 0.3419 & $+0.0000$ \\
  \cmidrule(l){2-13}
  & \multirow{7}{*}{\shortstack{None \\ (unfiltered)}}
  & 10 & 10 & 17 & SVR & 0.4323 & 0.4708 & 9  & SVR & 0.4685 & 0.4525 & $-0.0362$ \\
  & & 20 & 20 & 18 & SVR & 0.4723 & 0.4488 & 9  & SVR & 0.4681 & 0.4664 & $+0.0042$ \\
  & & 30 & 30 & 8  & SVR & 0.5009 & 0.4458 & 17 & SVR & 0.4919 & 0.4457 & $+0.0090$ \\
  & & 40 & 38 & 18 & SVR & 0.5211 & 0.4292 & 18 & SVR & 0.5211 & 0.4292 & $+0.0000$ \\
  & & 50 & 40 & 9  & SVR & 0.5214 & 0.4327 & 9  & SVR & 0.5214 & 0.4327 & $+0.0000$ \\
  & & 60 & 40 & 9  & SVR & 0.5214 & 0.4327 & 9  & SVR & 0.5214 & 0.4327 & $+0.0000$ \\
  & & 72 & 40 & 9  & SVR & 0.5214 & 0.4327 & 9  & SVR & 0.5214 & 0.4327 & $+0.0000$ \\
  \midrule
  \multirow{7}{*}{*N$_2$H}
  & \multirow{7}{*}{\shortstack{$|z|\leq3$ \\ (outlier-removed)}}
  & 10 & 10 & 15 & XGB & 0.7866 & 0.3974 & 3  & XGB & 0.7878 & 0.3757 & $-0.0012$ \\
  & & 20 & 20 & 2  & XGB & 0.7975 & 0.3760 & 3  & XGB & 0.7957 & 0.3692 & $+0.0018$ \\
  & & 30 & 30 & 17 & XGB & 0.7985 & 0.3813 & 3  & GB  & 0.7959 & 0.3951 & $+0.0026$ \\
  & & 40 & 35 & 17 & XGB & 0.7960 & 0.3816 & 17 & XGB & 0.7960 & 0.3816 & $+0.0000$ \\
  & & 50 & 35 & 17 & XGB & 0.7960 & 0.3816 & 17 & XGB & 0.7960 & 0.3816 & $+0.0000$ \\
  & & 60 & 35 & 17 & XGB & 0.7960 & 0.3816 & 17 & XGB & 0.7960 & 0.3816 & $+0.0000$ \\
  & & 72 & 35 & 17 & XGB & 0.7960 & 0.3816 & 17 & XGB & 0.7960 & 0.3816 & $+0.0000$ \\
  \cmidrule(l){2-13}
  & \multirow{7}{*}{\shortstack{None \\ (unfiltered)}}
  & 10 & 10 & 14 & SVR & 0.6827 & 0.4468 & 4  & SVR & 0.6795 & 0.4318 & $+0.0032$ \\
  & & 20 & 20 & 2  & XGB & 0.6890 & 0.4468 & 3  & GB  & 0.6782 & 0.4560 & $+0.0108$ \\
  & & 30 & 30 & 2  & XGB & 0.7087 & 0.4307 & 19 & ExtraTrees & 0.6930 & 0.4378 & $+0.0157$ \\
  & & 40 & 39 & 3  & XGB & 0.7086 & 0.4216 & 3  & XGB & 0.7086 & 0.4216 & $+0.0000$ \\
  & & 50 & 39 & 3  & XGB & 0.7086 & 0.4216 & 3  & XGB & 0.7086 & 0.4216 & $+0.0000$ \\
  & & 60 & 39 & 3  & XGB & 0.7086 & 0.4216 & 3  & XGB & 0.7086 & 0.4216 & $+0.0000$ \\
  & & 72 & 39 & 3  & XGB & 0.7086 & 0.4216 & 3  & XGB & 0.7086 & 0.4216 & $+0.0000$ \\
  \midrule
  \multirow{7}{*}{*NH$_3$}
  & \multirow{7}{*}{\shortstack{$|z|\leq3$ \\ (outlier-removed)}}
  & 10 & 10 & 20 & XGB & 0.6328 & 0.1971 & 19 & XGB & 0.6578 & 0.1866 & $-0.0250$ \\
  & & 20 & 20 & 20 & XGB & 0.6590 & 0.1900 & 19 & XGB & 0.6971 & 0.1769 & $-0.0380$ \\
  & & 30 & 30 & 19 & XGB & 0.6731 & 0.1857 & 19 & XGB & 0.6926 & 0.1778 & $-0.0195$ \\
  & & 40 & 37 & 19 & XGB & 0.6718 & 0.1874 & 19 & XGB & 0.6718 & 0.1874 & $+0.0000$ \\
  & & 50 & 37 & 19 & XGB & 0.6718 & 0.1874 & 19 & XGB & 0.6718 & 0.1874 & $+0.0000$ \\
  & & 60 & 37 & 19 & XGB & 0.6718 & 0.1874 & 19 & XGB & 0.6718 & 0.1874 & $+0.0000$ \\
  & & 72 & 37 & 19 & XGB & 0.6718 & 0.1874 & 19 & XGB & 0.6718 & 0.1874 & $+0.0000$ \\
  \cmidrule(l){2-13}
  & \multirow{7}{*}{\shortstack{None \\ (unfiltered)}}
  & 10 & 10 & 19 & XGB & 0.5884 & 0.2440 & 20 & XGB & 0.5828 & 0.2409 & $+0.0056$ \\
  & & 20 & 20 & 20 & XGB & 0.6579 & 0.2312 & 19 & XGB & 0.6409 & 0.2283 & $+0.0170$ \\
  & & 30 & 30 & 20 & XGB & 0.6733 & 0.2281 & 19 & XGB & 0.6699 & 0.2211 & $+0.0033$ \\
  & & 40 & 37 & 19 & XGB & 0.6692 & 0.2221 & 19 & XGB & 0.6692 & 0.2221 & $+0.0000$ \\
  & & 50 & 37 & 19 & XGB & 0.6692 & 0.2221 & 19 & XGB & 0.6692 & 0.2221 & $+0.0000$ \\
  & & 60 & 37 & 19 & XGB & 0.6692 & 0.2221 & 19 & XGB & 0.6692 & 0.2221 & $+0.0000$ \\
  & & 72 & 37 & 19 & XGB & 0.6692 & 0.2221 & 19 & XGB & 0.6692 & 0.2221 & $+0.0000$ \\
\end{longtable}
\normalsize
\end{landscape}
\doublespacing

\clearpage

\singlespacing
\small
\setlength{\LTcapwidth}{\linewidth}
\begin{longtable}{lcccrrcc}
  \caption{Summary statistics of N$_2$ adsorption across intermetallic alloys (IMCs).
  Reported values include the minimum and maximum adsorption energies per alloy, the number
  of sites for molecular versus atomic adsorption, and the orientation of molecularly adsorbed
  N$_2$ (side-on, or end-on adsorption).}
  \label{tab:tabS6}\\
  \toprule
  \textbf{Material} & \textbf{Total} & \textbf{No.\ site} & \textbf{No.\ site} &
  \textbf{Min} & \textbf{Max} & \textbf{No.\ end-on} & \textbf{No.\ side-on} \\
   & \textbf{No.\ sites} & \textbf{(mol.\ ads)} & \textbf{(at.\ ads)} &
  \textbf{$E_\mathrm{ads}$} & \textbf{$E_\mathrm{ads}$} & \textbf{ads} & \textbf{ads} \\
  \midrule
  \endfirsthead
  \toprule
  \textbf{Material} & \textbf{Total} & \textbf{No.\ site} & \textbf{No.\ site} &
  \textbf{Min} & \textbf{Max} & \textbf{No.\ end-on} & \textbf{No.\ side-on} \\
   & \textbf{No.\ sites} & \textbf{(mol.\ ads)} & \textbf{(at.\ ads)} &
  \textbf{$E_\mathrm{ads}$} & \textbf{$E_\mathrm{ads}$} & \textbf{ads} & \textbf{ads} \\
  \midrule
  \endhead
  \bottomrule
  \endfoot
  Co$_3$Ni     & 7  & 7  & 0  & $0.139$  & $0.406$  & 0 & 7  \\
  Al$_3$Ni$_2$ & 8  & 2  & 5  & $-0.249$ & $0.288$  & 1 & 1  \\
  Zn$_{11}$Co$_2$ & 10 & 2 & 0 & $0.011$ & $0.359$ & 0 & 2  \\
  Al$_3$V      & 7  & 3  & 2  & $-0.854$ & $-0.020$ & 0 & 3  \\
  AlFe         & 10 & 8  & 1  & $-0.888$ & $0.425$  & 3 & 5  \\
  Zn$_3$Cu     & 7  & 1  & 0  & $0.034$  & $2.418$  & 1 & 0  \\
  Zn$_8$Cu$_5$ & 4  & 1  & 3  & $1.149$  & $3.913$  & 0 & 1  \\
  AlV$_3$      & 13 & 3  & 10 & $-3.360$ & $-1.837$ & 0 & 3  \\
  ZnCu         & 5  & 0  & 1  & $0.061$  & $2.687$  & 0 & 0  \\
  V$_3$Ni$_2$  & 19 & 15 & 3  & $-2.646$ & $-0.339$ & 4 & 11 \\
  Co$_3$Pt     & 4  & 3  & 0  & $-0.182$ & $2.970$  & 1 & 2  \\
  Al$_2$Cu     & 5  & 2  & 1  & $-2.856$ & $-1.538$ & 2 & 0  \\
  Cu$_3$Pt     & 7  & 1  & 0  & $0.011$  & $2.084$  & 0 & 1  \\
  CuPt$_7$     & 7  & 6  & 1  & $-0.303$ & $3.201$  & 3 & 3  \\
  Al$_3$Ni     & 10 & 10 & 0  & $-0.554$ & $0.349$  & 5 & 5  \\
  CoPt$_3$     & 3  & 2  & 0  & $0.279$  & $1.395$  & 0 & 2  \\
  Al$_4$Cu$_9$ & 8  & 1  & 0  & $-0.317$ & $0.002$  & 1 & 0  \\
  Fe$_9$Co$_7$ & 6  & 3  & 1  & $-2.089$ & $0.403$  & 0 & 3  \\
  AlCo         & 7  & 6  & 0  & $-1.043$ & $0.030$  & 1 & 5  \\
  FeNi         & 6  & 5  & 0  & $1.099$  & $4.420$  & 0 & 5  \\
  AlFe$_3$     & 9  & 8  & 0  & $-0.637$ & $0.211$  & 2 & 6  \\
  Ni$_3$Pt     & 6  & 6  & 0  & $0.277$  & $1.260$  & 0 & 6  \\
  NiPt$_3$     & 4  & 3  & 0  & $-0.179$ & $0.579$  & 0 & 3  \\
  V$_3$Co      & 16 & 8  & 8  & $-2.128$ & $-0.314$ & 0 & 8  \\
  V$_3$Pt      & 6  & 2  & 4  & $-3.733$ & $-0.516$ & 0 & 2  \\
  VNi$_3$      & 15 & 14 & 1  & $-0.455$ & $2.319$  & 0 & 14 \\
  Al$_2$Pt     & 7  & 0  & 1  & $-0.027$ & $2.152$  & 0 & 0  \\
  Al$_3$Pt$_2$ & 8  & 2  & 2  & $-0.250$ & $1.292$  & 1 & 1  \\
  AlNi         & 7  & 4  & 1  & $-0.097$ & $1.915$  & 1 & 3  \\
  Zn$_3$Pt     & 11 & 0  & 1  & $0.049$  & $3.386$  & 0 & 0  \\
  AlPt$_3$     & 14 & 10 & 2  & $-0.548$ & $3.162$  & 2 & 8  \\
  Zn$_9$Fe$_4$ & 10 & 1  & 0  & $0.015$  & $0.411$  & 0 & 1  \\
  V$_3$Fe      & 16 & 10 & 6  & $-2.300$ & $-1.241$ & 0 & 10 \\
  ZnNi         & 6  & 6  & 0  & $-0.522$ & $0.998$  & 2 & 4  \\
  VCo$_3$      & 9  & 8  & 0  & $-0.732$ & $-0.035$ & 2 & 6  \\
  ZnPt$_3$     & 6  & 5  & 0  & $0.092$  & $0.505$  & 3 & 2  \\
  VZn$_3$      & 8  & 7  & 1  & $-1.521$ & $-0.758$ & 4 & 3  \\
  Fe$_3$Co     & 6  & 4  & 2  & $-0.528$ & $3.671$  & 0 & 4  \\
  FeCo         & 5  & 4  & 1  & $-0.388$ & $4.587$  & 0 & 4  \\
  AlNi$_3$     & 9  & 7  & 1  & $-0.341$ & $2.026$  & 3 & 4  \\
  FeNi$_3$     & 7  & 5  & 1  & $0.270$  & $4.082$  & 0 & 5  \\
  AlV          & 12 & 8  & 4  & $-3.527$ & $-0.626$ & 1 & 7  \\
  FePt$_3$     & 6  & 3  & 0  & $0.120$  & $1.638$  & 0 & 3  \\
  NiPt         & 4  & 4  & 0  & $-0.988$ & $0.072$  & 0 & 4  \\
  V$_3$Ni      & 15 & 7  & 8  & $-2.258$ & $-0.919$ & 0 & 7  \\
  VFe$_3$      & 11 & 10 & 0  & $-1.214$ & $0.135$  & 0 & 10 \\
  VPt          & 1  & 0  & 0  & $0.134$  & $0.134$  & 0 & 0  \\
\end{longtable}
\normalsize
\doublespacing
\clearpage

\singlespacing
\begin{longtable}{lcccccc}
  \caption{Summary statistics of N$_2$H adsorption across intermetallic alloys (IMCs).
  Reported values include the minimum and maximum adsorption energies per alloy, the number
  of sites for molecular adsorption, and the orientation of molecularly adsorbed
  N$_2$H (side-on, or end-on adsorption).}
  \label{tab:tabS7}\\
  \toprule
  \textbf{Material} & \textbf{Total} & \textbf{No.\ site} & \textbf{Min} & \textbf{Max} &
  \textbf{No.\ end-on} & \textbf{No.\ side-on} \\
   & \textbf{No.\ sites} & \textbf{(mol.\ ads)} &
  \textbf{$E_\mathrm{ads}$} & \textbf{$E_\mathrm{ads}$} & \textbf{ads} & \textbf{ads} \\
  \midrule
  \endfirsthead
  \toprule
  \textbf{Material} & \textbf{Total} & \textbf{No.\ site} & \textbf{Min} & \textbf{Max} &
  \textbf{No.\ end-on} & \textbf{No.\ side-on} \\
   & \textbf{No.\ sites} & \textbf{(mol.\ ads)} &
  \textbf{$E_\mathrm{ads}$} & \textbf{$E_\mathrm{ads}$} & \textbf{ads} & \textbf{ads} \\
  \midrule
  \endhead
  \bottomrule
  \endfoot
  Co$_3$Ni        & 8  & 0  & $-3.526$ & $4.570$  & 0 & 0  \\
  Al$_3$Ni$_2$    & 8  & 4  & $-3.065$ & $-2.239$ & 0 & 4  \\
  Zn$_{11}$Co$_2$ & 11 & 10 & $-2.609$ & $1.088$  & 0 & 10 \\
  Al$_3$V         & 15 & 13 & $-4.184$ & $-2.187$ & 1 & 12 \\
  AlFe            & 10 & 8  & $-4.502$ & $-2.355$ & 0 & 8  \\
  Zn$_3$Cu        & 8  & 5  & $-1.314$ & $2.287$  & 1 & 4  \\
  Zn$_8$Cu$_5$    & 6  & 0  & $-1.308$ & $4.483$  & 0 & 0  \\
  AlV$_3$         & 2  & 2  & $-4.978$ & $-4.814$ & 0 & 2  \\
  ZnCu            & 7  & 3  & $-1.225$ & $3.240$  & 0 & 3  \\
  V$_3$Ni$_2$     & 16 & 12 & $-4.999$ & $-2.378$ & 3 & 9  \\
  Co$_3$Pt        & 8  & 2  & $-2.721$ & $3.974$  & 0 & 2  \\
  Al$_2$Cu        & 2  & 2  & $-4.691$ & $-3.806$ & 0 & 2  \\
  Cu$_3$Pt        & 6  & 1  & $-0.425$ & $3.407$  & 0 & 1  \\
  CuPt$_7$        & 8  & 5  & $-2.522$ & $4.989$  & 0 & 5  \\
  Al$_3$Ni        & 9  & 6  & $-4.375$ & $-1.685$ & 0 & 6  \\
  CoPt$_3$        & 6  & 5  & $-1.739$ & $0.499$  & 0 & 5  \\
  Al$_4$Cu$_9$    & 8  & 7  & $-1.620$ & $1.384$  & 0 & 7  \\
  Fe$_9$Co$_7$    & 5  & 3  & $-3.856$ & $2.045$  & 2 & 1  \\
  AlCo            & 7  & 7  & $-3.220$ & $-2.476$ & 0 & 7  \\
  FeNi            & 6  & 4  & $-2.719$ & $1.394$  & 0 & 4  \\
  AlFe$_3$        & 9  & 7  & $-4.325$ & $-2.091$ & 0 & 7  \\
  Ni$_3$Pt        & 5  & 1  & $-1.818$ & $3.272$  & 0 & 1  \\
  NiPt$_3$        & 5  & 3  & $-2.353$ & $4.830$  & 0 & 3  \\
  V$_3$Co         & 14 & 14 & $-4.545$ & $-3.363$ & 0 & 14 \\
  V$_3$Pt         & 6  & 2  & $-4.064$ & $1.811$  & 0 & 2  \\
  VNi$_3$         & 16 & 12 & $-4.849$ & $-1.806$ & 1 & 11 \\
  Al$_2$Pt        & 7  & 7  & $-2.401$ & $-1.488$ & 1 & 6  \\
  Al$_3$Pt$_2$    & 8  & 7  & $-3.528$ & $-1.833$ & 0 & 7  \\
  AlNi            & 7  & 7  & $-2.273$ & $-1.435$ & 1 & 6  \\
  Zn$_3$Pt        & 13 & 4  & $-0.840$ & $4.795$  & 0 & 4  \\
  AlPt$_3$        & 15 & 14 & $-2.642$ & $-1.019$ & 2 & 12 \\
  Zn$_9$Fe$_4$    & 10 & 9  & $-2.070$ & $-1.081$ & 0 & 9  \\
  V$_3$Fe         & 15 & 14 & $-4.506$ & $-3.858$ & 0 & 14 \\
  ZnNi            & 7  & 2  & $-2.190$ & $4.483$  & 0 & 2  \\
  VCo$_3$         & 9  & 6  & $-4.474$ & $-3.099$ & 0 & 6  \\
  ZnPt$_3$        & 4  & 3  & $-1.737$ & $-0.967$ & 0 & 3  \\
  VZn$_3$         & 8  & 6  & $-4.162$ & $-0.059$ & 0 & 6  \\
  Fe$_3$Co        & 4  & 2  & $-3.898$ & $4.653$  & 0 & 2  \\
  FeCo            & 4  & 1  & $-3.811$ & $4.149$  & 0 & 1  \\
  AlNi$_3$        & 8  & 6  & $-2.311$ & $-0.870$ & 0 & 6  \\
  FeNi$_3$        & 9  & 3  & $-2.815$ & $4.849$  & 0 & 3  \\
  AlV             & 11 & 9  & $-4.433$ & $-3.212$ & 0 & 9  \\
  FePt$_3$        & 2  & 2  & $-0.362$ & $0.458$  & 0 & 2  \\
  NiPt            & 3  & 1  & $-4.539$ & $-2.867$ & 0 & 1  \\
  V$_3$Ni         & 14 & 13 & $-4.600$ & $-3.404$ & 0 & 13 \\
  VFe$_3$         & 8  & 8  & $-3.856$ & $-2.937$ & 0 & 8  \\
  VPt             & 3  & 2  & $-1.635$ & $0.060$  & 0 & 2  \\
\end{longtable}
\doublespacing

\clearpage

\singlespacing
\begin{longtable}{lcccc}
  \caption{Summary statistics of NH$_3$ adsorption across intermetallic alloys (IMCs). The
  reported values include the minimum and maximum adsorption energies among all surface sites
  for each alloy, along with the number of sites corresponding to molecular
  adsorption.}
  \label{tab:tabS8}\\
  \toprule
  \textbf{Material} & \textbf{Total No.\ sites} & \textbf{No.\ site (mol.\ ads)} &
  \textbf{Min $E_\mathrm{ads}$} & \textbf{Max $E_\mathrm{ads}$} \\
  \midrule
  \endfirsthead
  \toprule
  \textbf{Material} & \textbf{Total No.\ sites} & \textbf{No.\ site (mol.\ ads)} &
  \textbf{Min $E_\mathrm{ads}$} & \textbf{Max $E_\mathrm{ads}$} \\
  \midrule
  \endhead
  \bottomrule
  \endfoot
  Co$_3$Ni        & 7  & 4  & $-0.783$ & $-0.261$ \\
  Al$_3$Ni$_2$    & 9  & 5  & $-1.070$ & $-0.078$ \\
  Zn$_{11}$Co$_2$ & 11 & 8  & $-1.331$ & $3.422$  \\
  Al$_3$V         & 16 & 15 & $-1.650$ & $-0.224$ \\
  AlFe            & 10 & 9  & $-1.257$ & $-0.214$ \\
  Zn$_3$Cu        & 7  & 4  & $-0.251$ & $-0.003$ \\
  Zn$_8$Cu$_5$    & 5  & 0  & $0.653$  & $4.776$  \\
  AlV$_3$         & 13 & 12 & $-2.454$ & $4.242$  \\
  ZnCu            & 7  & 5  & $-0.520$ & $3.410$  \\
  V$_3$Ni$_2$     & 21 & 19 & $-3.881$ & $-0.934$ \\
  Co$_3$Pt        & 9  & 3  & $-0.668$ & $3.836$  \\
  Al$_2$Cu        & 5  & 5  & $-3.782$ & $-2.901$ \\
  Cu$_3$Pt        & 8  & 2  & $-0.385$ & $3.423$  \\
  CuPt$_7$        & 9  & 4  & $-1.140$ & $2.557$  \\
  Al$_3$Ni        & 10 & 9  & $-0.841$ & $-0.088$ \\
  CoPt$_3$        & 8  & 3  & $-0.801$ & $2.306$  \\
  Al$_4$Cu$_9$    & 7  & 7  & $-1.240$ & $-0.398$ \\
  Fe$_9$Co$_7$    & 6  & 5  & $-1.037$ & $-0.507$ \\
  AlCo            & 7  & 7  & $-1.005$ & $-0.825$ \\
  FeNi            & 8  & 5  & $0.438$  & $1.878$  \\
  AlFe$_3$        & 8  & 6  & $-0.721$ & $1.476$  \\
  Ni$_3$Pt        & 7  & 6  & $-0.586$ & $0.084$  \\
  NiPt$_3$        & 8  & 3  & $-1.227$ & $1.835$  \\
  V$_3$Co         & 16 & 15 & $-1.397$ & $-0.726$ \\
  V$_3$Pt         & 9  & 7  & $-1.416$ & $4.849$  \\
  VNi$_3$         & 16 & 14 & $-1.295$ & $1.955$  \\
  Al$_2$Pt        & 7  & 5  & $-0.926$ & $-0.042$ \\
  Al$_3$Pt$_2$    & 7  & 4  & $-1.114$ & $-0.067$ \\
  AlNi            & 7  & 6  & $-0.808$ & $-0.054$ \\
  Zn$_3$Pt        & 15 & 10 & $-0.364$ & $4.685$  \\
  AlPt$_3$        & 13 & 13 & $-1.260$ & $-0.954$ \\
  Zn$_9$Fe$_4$    & 10 & 9  & $-0.678$ & $4.666$  \\
  V$_3$Fe         & 16 & 15 & $-2.409$ & $-0.743$ \\
  ZnNi            & 10 & 9  & $-0.761$ & $2.435$  \\
  VCo$_3$         & 9  & 6  & $-1.186$ & $0.237$  \\
  ZnPt$_3$        & 8  & 4  & $-0.707$ & $3.422$  \\
  VZn$_3$         & 9  & 6  & $-1.782$ & $-0.267$ \\
  Fe$_3$Co        & 5  & 3  & $-0.761$ & $-0.085$ \\
  FeCo            & 6  & 4  & $-0.761$ & $2.865$  \\
  AlNi$_3$        & 8  & 6  & $-0.671$ & $2.074$  \\
  FeNi$_3$        & 7  & 5  & $-0.777$ & $-0.085$ \\
  AlV             & 12 & 12 & $-1.638$ & $-0.549$ \\
  FePt$_3$        & 7  & 3  & $-0.863$ & $-0.002$ \\
  NiPt            & 5  & 4  & $-3.276$ & $-0.086$ \\
  V$_3$Ni         & 16 & 14 & $-2.594$ & $-1.488$ \\
  VFe$_3$         & 11 & 11 & $-1.029$ & $-0.670$ \\
  VPt             & 4  & 0  & $-0.042$ & $4.369$  \\
\end{longtable}
\doublespacing

\clearpage

\singlespacing
\setlength{\LTcapwidth}{\linewidth}
\begin{longtable}{cp{8cm}cc}
  \caption{Mean absolute SHAP values for all 38 features used to train the best-performing
  model (XGBRegressor) for *N$_2$ adsorption energy prediction under the outlier-removed
  ($|z| \leq 3$) condition with a feature-selection budget of $K = 40$. Values are computed
  over the sealed test set.}
  \label{tab:tabS9}\\
  \toprule
  \textbf{Rank} & \textbf{Feature} & \textbf{Symbol} & \textbf{Mean $|\mathrm{SHAP}|$} \\
  \midrule
  \endfirsthead
  \toprule
  \textbf{Rank} & \textbf{Feature} & \textbf{Symbol} & \textbf{Mean $|\mathrm{SHAP}|$} \\
  \midrule
  \endhead
  \bottomrule
  \endfoot
  1  & Maximum covalent radius in the neighbor shell & $\max(r_{\mathrm{cov,shell}})$ & 0.284 \\
  2  & Mean electronegativity of the shell & $\mu(\chi_{\mathrm{shell}})$ & 0.109 \\
  3  & Standard deviation of electron affinity in the shell & $\sigma(\mathrm{EA}_{\mathrm{shell}})$ & 0.053 \\
  4  & Bader charge of the first nearest atom to adsorbate & $q_0$ & 0.050 \\
  5  & Mean ionization energy of the shell & $\mu(\mathrm{IE}_{\mathrm{shell}})$ & 0.049 \\
  6  & Local p-band filling & $f_{p,\mathrm{local}}$ & 0.047 \\
  7  & Local s-band center & $\varepsilon_{s,\mathrm{local}}$ & 0.039 \\
  8  & Distance of the first nearest atom from the adsorbate & $d_0$ & 0.033 \\
  9  & Total s-band filling & $f_{s,\mathrm{total}}$ & 0.028 \\
  10 & Local d-band filling & $f_{d,\mathrm{local}}$ & 0.023 \\
  11 & Local d-band center & $\varepsilon_{d,\mathrm{local}}$ & 0.022 \\
  12 & Mean atomic number of the shell & $\mu(Z_{\mathrm{shell}})$ & 0.022 \\
  13 & Covalent radius of the first nearest atom to adsorbate & $r_{0,\mathrm{cov}}$ & 0.019 \\
  14 & Standard deviation of ionization energy in the shell & $\sigma(\mathrm{IE}_{\mathrm{shell}})$ & 0.015 \\
  15 & Total p-band filling & $f_{p,\mathrm{total}}$ & 0.014 \\
  16 & Mean electron affinity of the shell & $\mu(\mathrm{EA}_{\mathrm{shell}})$ & 0.013 \\
  17 & Standard deviation of Bader charge in the shell & $\sigma(q_{\mathrm{shell}})$ & 0.013 \\
  18 & Maximum Bader charge in the shell & $\max(q_{\mathrm{shell}})$ & 0.009 \\
  19 & Crystal type & --- & 0.009 \\
  20 & Local p-band center & $\varepsilon_{p,\mathrm{local}}$ & 0.008 \\
  21 & Standard deviation of atomic number in the shell & $\sigma(Z_{\mathrm{shell}})$ & 0.008 \\
  22 & Local s-band filling & $f_{s,\mathrm{local}}$ & 0.005 \\
  23 & Minimum electronegativity in the shell & $\min(\chi_{\mathrm{shell}})$ & 0.004 \\
  24 & Maximum atomic number in the shell & $\max(Z_{\mathrm{shell}})$ & 0.004 \\
  25 & Atomic number of the first nearest atom to adsorbate & $Z_0$ & 0.003 \\
  26 & Maximum electronegativity in the shell & $\max(\chi_{\mathrm{shell}})$ & 0.002 \\
  27 & Space group & --- & 0.001 \\
  28 & Total d-band filling & $f_{d,\mathrm{total}}$ & 0.000 \\
  29 & First ionization energy of first nearest atom to adsorbate & $\mathrm{IE}_0$ & 0.000 \\
  30 & Electron affinity of the first nearest atom to adsorbate & $\mathrm{EA}_0$ & 0.000 \\
  31 & Maximum electron affinity in the shell & $\max(\mathrm{EA}_{\mathrm{shell}})$ & 0.000 \\
  32 & Minimum electron affinity in the shell & $\min(\mathrm{EA}_{\mathrm{shell}})$ & 0.000 \\
  33 & Minimum atomic number in the shell & $\min(Z_{\mathrm{shell}})$ & 0.000 \\
  34 & Range of electronegativity in the shell & $\Delta(\chi_{\mathrm{shell}})$ & 0.000 \\
  35 & Electronegativity of the first nearest atom to adsorbate & $\chi_0$ & 0.000 \\
  36 & Maximum ionization energy in the shell & $\max(\mathrm{IE}_{\mathrm{shell}})$ & 0.000 \\
  37 & Minimum Bader charge in the shell & $\min(q_{\mathrm{shell}})$ & 0.000 \\
  38 & Minimum ionization energy in the shell & $\min(\mathrm{IE}_{\mathrm{shell}})$ & 0.000 \\
\end{longtable}
\doublespacing

\bibliographystyle{unsrtnat}
\bibliography{references}